\documentclass[opre,nonblindrev]{informs3aa} 
\usepackage{informs_set}
\newif\ifcomm
\commtrue

\usepackage{comment}
\usepackage{caption}
\usepackage{mathtools}
\usepackage{hyperref}
\usepackage{epsfig}
\usepackage{subcaption}
\usepackage[table,xcdraw]{xcolor}
\usepackage{bbm}
\usepackage{url}
\usepackage{algorithm}
\usepackage{times}
\usepackage{amsmath, graphicx, latexsym, amssymb,  amsfonts}

\definecolor{teal}{rgb}{0.0, 0.5, 0.5}

\newcommand{\twopartdefnoif}[4]
{
	\left\{
		\begin{array}{ll}
			#1, & \quad #2, \\
			#3, & \quad #4
		\end{array}
	\right.
}

\newcommand{\threepartdefelse}[5]
{
	\left\{
		\begin{array}{lll}
			#1, & \quad\mbox{if } \quad #2, \\
			#3, & \quad\mbox{if } \quad #4, \\
			#5, & \quad\mbox{else }.
		\end{array}
	\right.
}

\ifcomm
    \newcounter{commentNumberI}
     \newcommand{\Kuang}[1]{\addtocounter{commentNumberI}{1}{{({\color{blue} {(\arabic{commentNumberI}.)} Kuang: #1})}}} 
       \newcommand{\Gal}[1]{\addtocounter{commentNumberI}{1}{{({\color{red} {(\arabic{commentNumberI}.)} Gal: #1})}}} 
    \newcommand{\A}[1]{\textbf{[#1]}}       
    
\else
    \newcommand{\A}[1]{}
    \newcommand{\Kuang}[1]{}
    \newcommand{\Gal}[1]{}
\fi
\newcommand{\kx}[1]{\Kuang{#1}}

\newcommand{\calE}{\mathcal{E}}

\newcommand{\calQ}{\mathcal{Q}}
\newcommand{\calD}{\mathcal{D}}
\newcommand{\calN}{\mathcal{N}}
\newcommand{\calX}{\mathcal{X}}
\newcommand{\zp}{\mathbb{Z}_+}
\newcommand{\rp}{\mathbb{R}_+}
\newcommand{\N}{\mathbb{N}}
\newcommand{\EE}{\mathbb{E}}
\newcommand{\cov}{\mbox{Cov}}
\newcommand{\var}{\mbox{Var}}
\newcommand{\nln}{\nonumber\\}
\newcommand{\pb}{\mathbb{P}}
\newcommand{\p}[1]{\left(#1 \right)}

\newcommand{\sk}[2]{\stackrel{#1}{#2}}
\newcommand{\lone}[1]{\left\| #1 \right\|}

\DoubleSpacedXI 

\usepackage{natbib}
 \bibpunct[, ]{(}{)}{,}{a}{}{,}%
 \def\bibfont{\small}%
\begin{document}


\RUNAUTHOR{Kuang and Mendelson}

\RUNTITLE{Learning Service Slowdown using Observational Data}

\TITLE{Learning Service Slowdown using Observational Data}

\ARTICLEAUTHORS{%
\AUTHOR{Xu Kuang}
\AFF{Graduate School of Business, Stanford University,  \EMAIL{kuangxu@stanford.edu}
\AUTHOR{Gal Mendelson
\footnote{This version: \today. G.~Mendelson was supported in part by a postdoctoral fellowship at Stanford University. } 
}
\AFF{Faculty of Data and Decision Sciences, Technion,  \EMAIL{galmen@technion.ac.il}} 
}
} 

\ABSTRACT{ 
Being able to identify service slowdowns is crucial to many operational problems. We study how to use observational congestion data to learn service slowdown in a multi-server system that uses adaptive congestion control mechanisms. 
We show that a commonly used summary statistic that relies on the marginal congestion measured at individual servers can be highly inaccurate in the presence of adaptive congestion control. We propose a new statistic based on potential routing actions, and show it provides a much more robust signal for server slowdown in these settings. Unlike the marginal statistic, potential action aims to detect changes in the routing actions, and is able to uncover slowdowns even when they do not reflect in marginal congestion. Our results highlight the complexity in performing observational statistical analysis for service systems in the presence of adaptive congestion control. They also suggest that practitioners may want to combine multiple, orthogonal statistics to achieve reliable slowdown detection. 

\emph{Keywords: adaptive information erasure, congestion, observational data, summary statistic, slowdown detection. }
}

\maketitle

\section{Introduction}

Service systems power much of the modern economy, from supply chains, healthcare to data centers. At the heart of any service system lie the processing resources, often referred to as the {servers}. While their form can vary widely, from a doctor in an Emergency Department to a computer server in a data center, all servers share  a core functionality: to render service and satisfy demand in a consistent and reliable manner. It is therefore a system operator's foremost responsibility to be aware if a server experiences significant degradation in speed and processing capability and apply timely intervention. 

The main objective of this paper is to understand how to effectively learn service slowdown. Service slowdowns can be broadly divided into two categories depending on their visibility. In the first group, which we call \emph{overt slowdown}, the server experiences a major malfunction that can be flagged by a software instrumentation or reporting mechanism. A computer server that stops responding to health-checks, or a doctor who calls in sick both fall into this category, and learning slowdown in this setting is generally simple due to the overt nature of the failures. In the second group, which we call \emph{covert slowdown}, the server's processing speed decreases but the degradation is so minor or gradual that it does not set off an alarm directly. Examples in this group include a virtual machine that slows down due to other competing workloads hosted on the physical server, which is not immediately visible to the user of the said virtual machine. Another example could be service slowdowns that occur at a healthcare facility as a result of gradual but persistent staff fatigue. 

While covert slowdowns may be less drastic than their overt counterparts, they nevertheless have a significant impact on operations. 
In the setting of an Emergency Department, even relatively mild decreases in processing rates can cause significant congestion, leading to increased wait times which are well known to have negative down-stream consequences on outcomes \citep{baker1991patients, derlet2002overcrowding}. Moreover, because covert slowdowns can be more difficult to detect, it may take longer for the operator to identify the root cause and apply mitigation, further amplifying the operational consequences of these events. 

In this paper, we study the learning of covert slowdowns in a service system using {observational congestion data}.  ``Observational'' in this context means that the data is collected during normal operations, as opposed to through deliberate randomized experiments. In most service systems, deliberate experiments are costly and seldom performed; observational data is therefore the more common source of information for performance monitoring. 
By ``congestion'' we are referring to the buildup of queues at the servers over time. Congestion is a natural information source, because it is often one of the first observable telltales of a reduced service capacity. Furthermore, state of congestion  is often already implemented as an input signal into a service system's congestion control mechanisms for dispatching, routing and capacity sizing. As a result, it is easily accessible without needing additional infrastructure or instrumentation. 

\begin{figure}[t!]
    \centering
    \includegraphics[scale=.55]{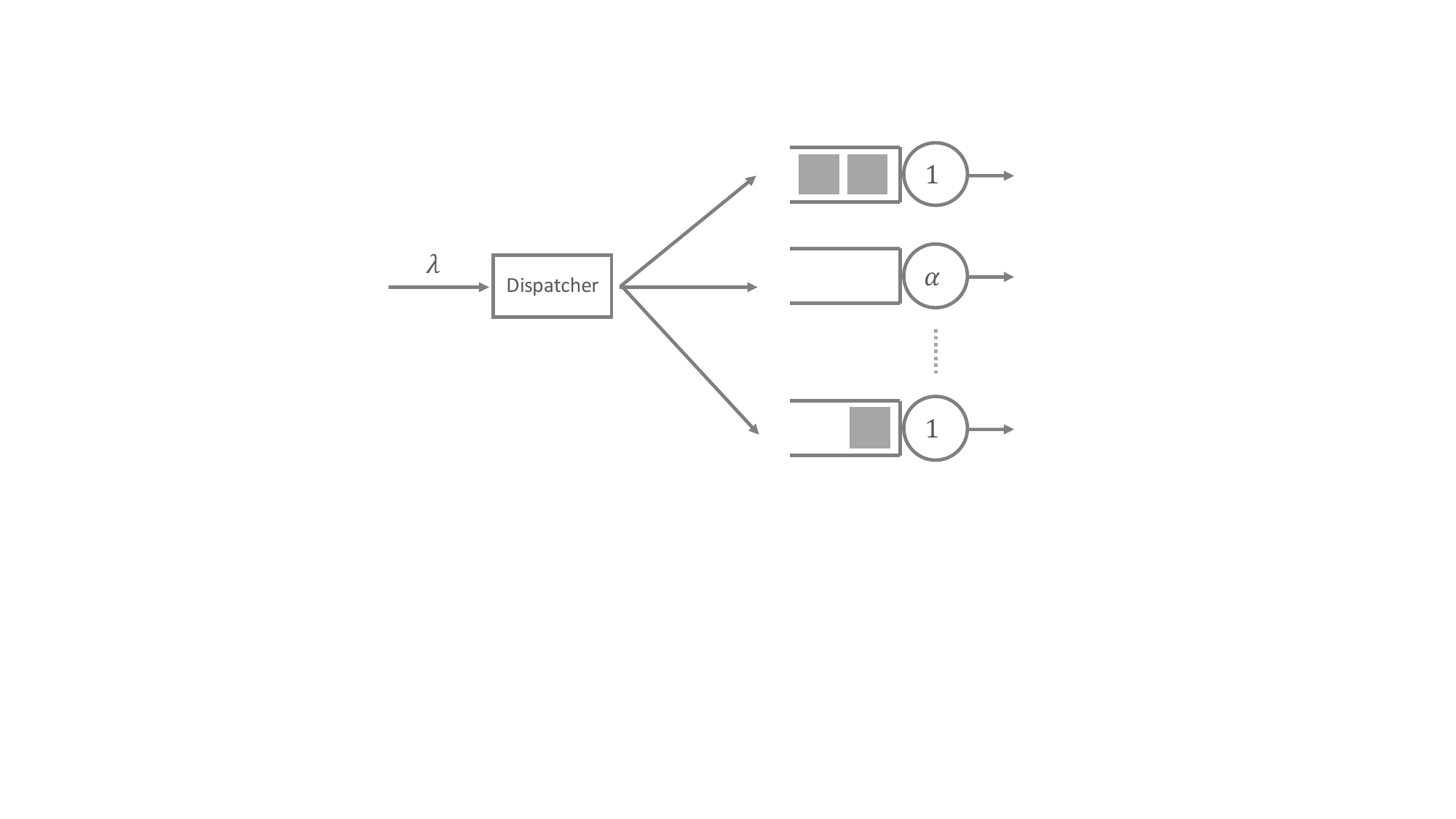}
    \caption{An illustration of the parallel multi-server system. Incoming jobs are sent to various servers by the dispatcher upon arrival. A slowdown refers to the event when the processing speed of the server, nominally at $\mu$, drops to $\alpha\mu$ for some $\alpha  \in (0,1)$. }
    \label{fig:sysdiag}
\end{figure}

A main insight of this work is that congestion information can be misleading when used for slowdown learning in a system with adaptive congestion control, and that a decision maker would benefit by utilizing statistics of actions, as well.  We begin by giving a description of the system model, followed by a preview of our key findings. We consider a parallel multi-server system, illustrated in Figure \ref{fig:sysdiag}. Incoming jobs are routed to one of $K$ servers by a {dispatcher}, who uses a congestion control mechanism to balance the incoming workload across different servers. A job then queues up at the designated server until a sufficient amount of service is rendered, at which point it departs from the system permanently. We will denote by $Q_i(t)$ the queue length at server $i$ and time $t$. Each server,  when healthy, is assumed to be rendering service unit rate. A slowdown refers to the event when the service rate of a server drops to $\alpha$ for some $0<\alpha<1$. 

\begin{figure}[t!]
    \centering
    \includegraphics[scale=.55]{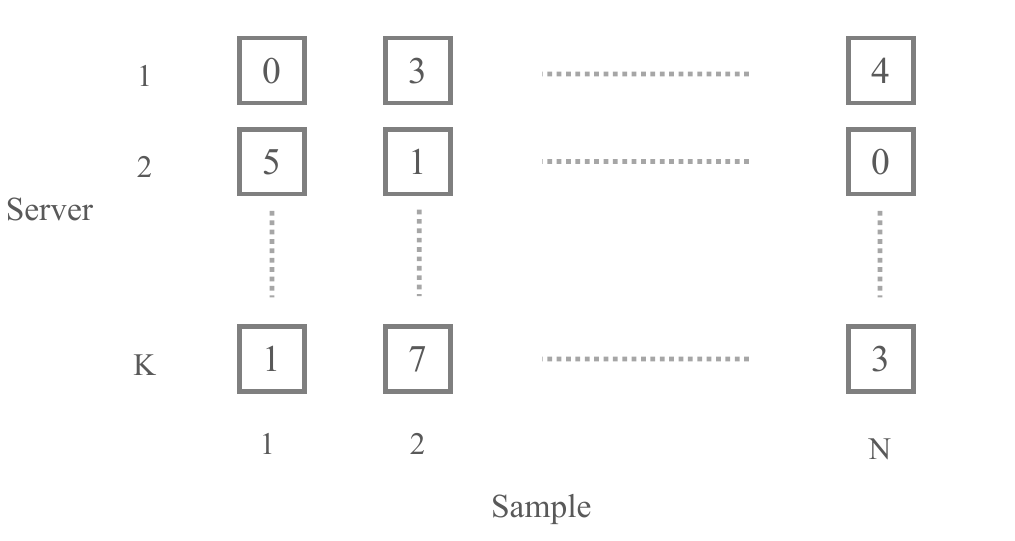}
    \caption{An illustration of the format of observational congestion data. Each column represents the queue lengths at the various servers at a particular point in time. The data set consists of $N$ such samples collected at different points in time. }
    \label{fig:datadiag}
\end{figure}

The decision maker has access to observational congestion data in a panel format, illustrated in Figure \ref{fig:datadiag}. The data consists of vector-valued queue length data sampled at $N$ distinct times, where each column represents the queue lengths at the various servers at the time of sampling: $(Q_1(t), Q_1(t), \ldots, Q_N(t))$. We will denote by $Q_{k,n}$, the queue length at server $k$ in the $n$th sample, and 
by $\calQ$ the the  $K\times N$ matrix panel of congestion data. 

The overall objective of the decision maker is to come up with an effective slowdown learning procedure that maps the panel $\calQ$ into a decision specifying whether a service slowdown has occurred, and if so, on which server. In practice, this process is often mediated via a {summary statistic}, which serves as a more succinct description of the data panel for easier interpretation. That is, the decision maker would first compute a summary statistic from $\calQ$, and henceforth use the value of such a statistic to perform detection. 

One of the most intuitive and commonly used summary statistic is {marginal (average) congestion}, which consists of the average queue length at each server $k$ across all $N$ samples, $\frac{1}{N}\sum_{n} Q_{k,n}$.  This marginal congestion statistic is a natural choice, as one would expect a server slowdown to cause its queue to build up. Using marginal congestion to learn slowdown has been popular: an algorithm would typically flag a server as experiencing a slowdown when its marginal congestion exceeds a pre-determined threshold, or is larger than the marginal congestion of other servers on a relative basis \citep[e.g.,][]{wang2011statistical, Yadav_2023}.

\subsection{Preview of Main Results} 

Our main research question is the following: How should the congestion data be analyzed, and specifically, what types of summary statistics should a decision maker use in order to effectively learn slowdown? Surprisingly, we find that  marginal congestion statistics 
can be highly uninformative when the dispatcher employs an adaptive congestion control scheme, such as the join-the-shortest-queue (JSQ) protocol. This is true even when the magnitude of the slowdown is substantial, and  the decision maker knows that such a slowdown has occurred somewhere in the system.  This negative result is alarming considering that adaptive congestion control mechanisms are widely deployed in service systems in both an explicit (e.g., leastconn algorithm in HAProxy \citep{HAproxyStarter2023}) and implicit manner (e.g., congestion-triggered ambulance diversion \citep{pham2006effects}). 

Specifically, we show that under JSQ, and, in fact, \textit{any} congestion control mechanism that achieves diffusion scale balance between the queues (often referred to as State-Space-Collapse, see \citep{stolyar2004maxweight}), any detection rule using the marginal congestion statistic would have a vanishing probability of detecting the slowed down server as the system become heavily loaded. Here, marginal congestion includes not only the average queue length statistic mentioned earlier, but an arbitrary function of the empirical marginal distribution of the queue length data, $(Q_{k,1}, Q_{k,2}, \ldots, Q_{k,N})$, subject to some regularity conditions.  

But why would marginal congestion, a common-sense summary statistic, fail? The root cause lies in what makes adaptive congestion control effective in the first place: its ability to equalize congestion across the system. When a slowdown occurs, the queue at the affected server tends to grow relative to those at healthy servers. However, this initial rise in marginal congestion is quickly counteracted by the congestion control mechanism, which adjusts dispatch decisions to send fewer new jobs to the slower server due to its elevated queue length. Over time, this feedback loop leads to marginal congestion levels between the slow and healthy servers becoming statistically indistinguishable. We term this phenomenon {adaptive information erasure}: the dispatcher responds to the initial congestion rise at the slow server by reducing the number of jobs sent there, thereby erasing evidence of the slowdown from the marginal congestion data. In essence, adaptive congestion control acts as a double-edged sword: while it helps mitigate local service-level fluctuations, it also removes critical information from marginal congestion statistics, undermining subsequent anomaly detection and statistical analysis that depend on it.

Motivated by the above insights, we propose a new class of summary statistics which we call {\bf potential action}, designed to robustly learn service slowdowns in the presence of adaptive congestion control. For each of the $N$ columns of the congestion data, $(Q_{1, n}, Q_{2, n}, \ldots, Q_{K,n})$, we generate a potential action, $A_{n}$, representing the index of the server to whom the dispatcher \textit{would have} sent a new job, if one were to arrive at this very moment. For instance, if the dispatcher uses a join-the-shortest-queue policy, then $A_{n}$ would be the index of the server with the shortest queue: $A_n = \argmin_k Q_{k,n}. $ (Assuming, for now, that such a server is unique.) The potential action statistic is defined to be the empirical distribution of the potential actions, $(A_1, A_2, \ldots, A_N)$. 

The potential action statistic is motivated by this key insight:  a capable adaptive congestion control, by design, reacts proportionally to the relative congestion levels across the servers.  When a server slows down, an adaptive congestion control mechanism would gradually send fewer jobs to the said server in the ensuing time periods, and this shift in turn will be reflected in the distributions of the potential actions. 

\begin{figure}[h]
    \centering
    \begin{subfigure}[t]{0.48\textwidth}
        \centering
        \includegraphics[width=\textwidth]{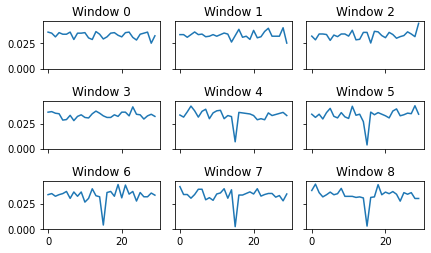}
        \caption{Potential Action}
        \label{fig:1a}
    \end{subfigure}
    \hfill
    \begin{subfigure}[t]{0.48\textwidth}
        \centering
        \includegraphics[width=\textwidth]{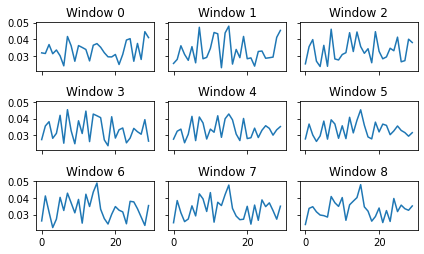}
        \caption{\small Marginal Congestion}
        \label{fig:1b}
    \end{subfigure}
    \caption{\small An illustration of potential action versus marginal congestion summary statistics when applied to the same congestion data measured across 9 consecutive time periods. The system contains 30 servers and a dispatcher using the join-the-shortest-queue policy, and  runs at a 95\% load. The x-axis is the identity of the server, and y-axis the value of the corresponding statistic. The  potential action statistic shows the empirical distribution of the potential actions across the servers, and marginal congestion the empirical average queue lengths. One can easily identify, using the potential action statistic, that server 15 experienced a slowdown starting from window 4, whereas it is very difficult to tell that from the marginal congestion statistic, which is considerably noisier and visually indistinguishable from one window to the next. 
    \label{fig:APtoMCcomparison}
  }
\end{figure}

 Concretely, we introduce a simple relative threshold test based on potential actions that carefully balances the risks of potentially expensive and service-disrupting false alarms (Type 1 error) versus missed detections (Type 2 error). We prove that if the possible slowdown magnitude is a priori known, and the samples in the congestion data are drawn independently from the system’s stationary distribution, the error -- defined as the maximum of Type 1 and Type 2 error probabilities -- decays exponentially with the sample size. We then extend the analysis to a more complex scenario where samples are drawn from the stationary distribution at equally spaced intervals, resulting in correlated samples. In this case, we prove that functions of the empirical averages of potential actions satisfy a Central Limit Theorem (CLT), provided the queue lengths evolve according to a geometrically ergodic Markov chain. As a special case, we prove this holds under JSQ. Leveraging the CLT, we derive a normal approximation for the error under our proposed test and demonstrate that it still approaches zero exponentially fast, as in the independent sample case. These results are robust as they hold across a wide range of maximally stable congestion control protocols, sufficiently heavy traffic intensities, and modest degrees of slowdowns. They stand in stark contrast to the failure of marginal statistics in these scenarios; Figure \ref{fig:APtoMCcomparison} visually illustrates this discrepancy.

We further provide a numerical study that complements our theoretical findings and offers additional evidence supporting the effectiveness of potential action-based learning procedures. Specifically, we demonstrate that the test error decays exponentially fast as a function of the sample size, with the rate of decay influenced by the number of servers, the slowdown factor, and the time interval between samples. 

Additionally, we address the scenario where the potential magnitude of the slowdown is unknown beforehand. We show that by selecting a minimal slowdown magnitude of interest and designing the test accordingly, comparable performance is achieved for more significant slowdowns. 

\begin{figure*}[h!]
    \centering
    \includegraphics[width=0.42\textwidth]{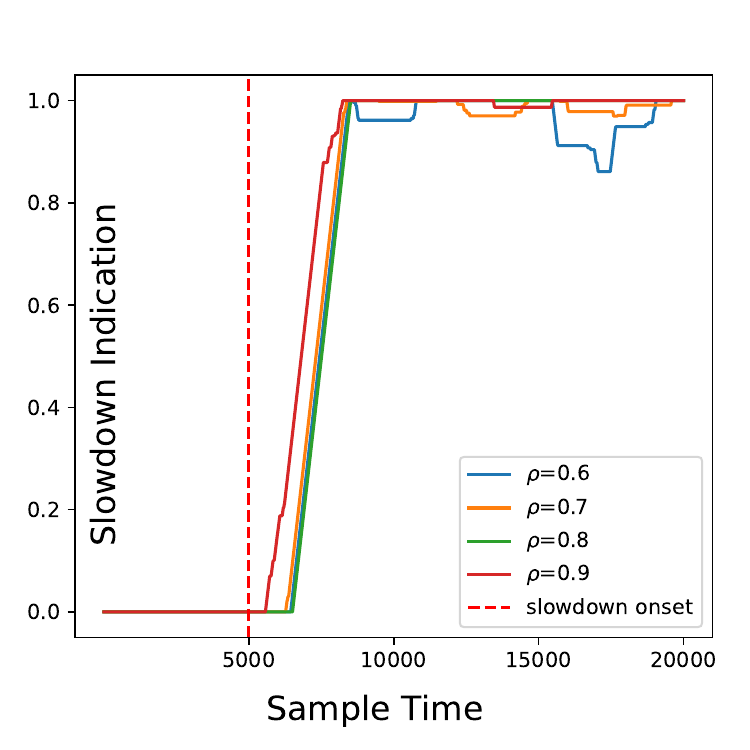}
    \caption{An operator's dashboard, indicating whether there is a slowdown (y-axis values around 1) in a system with 10 servers for different loads ($\rho$) and a slowdown of $60\%$ ($\alpha=0.4$) at sample time 5000.}
    \label{fig:dashboard_intro}
\end{figure*}

Finally, we propose a straightforward procedure capable of inferring slowdowns in an online manner with high probability. This illustrates how our results can be practically applied and displayed on a system operator’s dashboard for real-time monitoring. Figure \ref{fig:dashboard_intro} demonstrates how such a dashboard might appear, clearly indicating the presence of a slowdown in the system.

{\bf Potential Applications} The insights derived in this paper can potentially be applied to various service systems to enhance their reliability. For instance, modern data center anomaly detection primarily relies on machine learning algorithms that process a variety of input signals \citep{shirazi2017extended}. However, to our knowledge, potential action counts have not been utilized as input signals in load balancers like HAProxy. Our findings suggest that incorporating potential actions as an additional key metric could improve the effectiveness of these anomaly detection mechanisms.

A more complex application lies in Emergency Department (ED) congestion control. In regional hospital networks, overcrowded EDs often resort to ambulance diversion (AD) to manage congestion, redirecting incoming ambulances to less crowded facilities \citep{derlet2002overcrowding, pham2006effects, xu2016using}. Individual EDs can be viewed as servers, with service rates influenced by factors such as staffing levels and operational efficiency. While ambulance routing is typically not centrally coordinated, congestion is frequently cited as a key driver of AD decisions \citep{patel2006ambulance, pham2006effects}.

Detecting service slowdowns in EDs can support financial incentives aimed at promoting service excellence. For example, \cite{savva2019can} propose linking payments to congestion levels and wait times. However, because ambulance diversion tends to balance congestion across EDs, marginal congestion alone may not reliably indicate service quality. Well-run facilities may appear disadvantaged due to receiving diverted patients. Our findings suggest that action statistics—such as the frequency of diversion or the number of diverted patients received—provide a more comprehensive measure of relative performance in interconnected ED networks.


\subsection{Literature Review}
\label{sec:lit_review}

{\bf Anomaly Detection and Performance Monitoring} The present paper is motivated in part by congestion-based performance monitoring challenges that arise in data centers and other service systems. In a data center setting, marginal congestion statistics, such as the time-averaged buffer content sizes or connection counts, serve as as a crucial input signal for anomaly detection \citep[cf.~][]{ibidunmoye2015performance, shirazi2017extended}. For instance, \cite{wang2011statistical} stress the importance of using simple, intuitive summary statistics, and describe procedures where an alarm is raised when a running-average of the marginal congestion at a given resource (e.g., CPU) exceeds a pre-determined threshold. Similar threshold-based detection rules are also implemented in currently deployed load-balancing frameworks \citep{HAproxyStarter2023}. In the context of healthcare, service rate deficiencies, such as those due to under-staffing, have been identified as one of the main congestion drivers at Emergency Departments \citep{hoot2008systematic}, prompting calls to tie marginal congestion measurements to payments as a way to incentivize hospitals to quickly identify and rectify service slowdowns \citep{savva2019reduce}.  A major departure of our work from the above literature is the emphasis on understanding the statistical impact of using adaptive congestion control mechanisms. Marginal congestion statistics are indeed a natural choice when performing hypothesis testing or estimation when the measurements from one resource are reasonably independent from another, but their statistical power can diminish in the presence of interactions between different servers, such as those mediated by adaptive congestion control. Our paper therefore provides a fresh perspective on the use of congestion data in performance monitoring, and suggests that the decision maker should consider incorporating the potential action statistic or its variants as an additional input for extra robustness. 

{\bf Estimation and Learning in Queueing Systems}  Our work  builds on the literature on parameter estimation and inference in queueing systems. The high-level objective of this literature is to use observed characteristics such as arrival counts, service completions and queue lengths  to estimate system primitives such as the arrival and service rates \citep{clarke1957maximum, larson1990queue}. The reader is referred to \citet{bhat1987statistical}, and more recently \citet{asanjarani2021survey}, for an excellent review and the references therein. Similar to our paper, the heavy-traffic regime is often used to derive theoretical guarantees for estimation procedures \citep{whitt2002stochastic}. There are however some notable differences. We focus on understanding the impact of adaptive control on the choice of summary statistics, whereas existing work tends to focus on systems with non-adaptive allocation rules. We aim at the more restricted goal of slowdown detection rather than general parameter estimation \citep{chen1994parameter}. While accurate parameter estimates can be used for detecting slowdown, it would require more samples and refined knowledge of the service discipline, and slowdown detection can often be accomplished in a data-efficient manner with minimal assumptions on the congestion control policy and service discipline, as we demonstrate. 


There is an emergent literature around active learning and experimentation in queueing systems. Learning scheduling and assignment rules in queues are investigated from a multi-armed bandit perspective in \cite{choudhury2021job, freund2022efficient, freund2023quantifying, krishnasamy2021learning, zhong2022learning}. Recently, \cite{li2023experimenting} study randomized control trials and switchback experiments in the presence of stochastic queueing congestion that arise in  admission control problems \citep{borgs2014optimal, stidham1985optimal, xu2016using}. The reader is referred to \citet{walton2021learning} for an overview of recently developments in learning in queueing networks. In contrast to this paper, most prior work in this area relies on active experimentation and randomization, while we focus on using passively generated observational data. 

{\bf  Causal Inference and Experimentation} Using observational data to draw causal insights has a rich history in the statistics and econometrics literature. Here, the treatment probabilities may not be uniform across different actions and can depend on the covariates of the units under consideration. A central insight from this literature is that naive estimators of treatment effects, such as differencing the average outcomes between treatment and control groups, can be biased due to the skews inherent in the treatment assignments (e.g., more educated individuals may be more likely to interact with an experimental technology), and also that such biases can be removed by properly weighing the average effects by the inverse of their treatment propensities  \citep{heckman2008econometric, hirano2003efficient, horvitz1952generalization, robins1994estimation, rosenbaum1983central}.  While our setup differs significantly from those in this literature, there are some intriguing  analogs. We can think of the queues as units, routing actions as treatments, congestion as outcomes, and the congestion control mechanism as the treatment assignment policy. Then in both cases, biases of naive estimators seem to arise from the use of a state-dependent treatment assignment rule, and the potential action summary statistics can be thought as a way to use treatment propensities to help reduce such biases. 

Also related to this work in spirit is a growing literature on causal inference in the presence of cross-unit interference \citep{arownow2017,bajari2021multiple, hudgens_2008, johari2022experimental, manski2013identification, munro2021treatment, wager2021experimenting}. A main driver of the failure of marginal congestion statistics in our setting can be traced to the interference between the servers. This in turn is a direct result of using an adaptive congestion control policy that  allocates a {shared} stream of arrivals across the servers. This aspect of our system echos some of the findings in this literature that highlight the importance of accounting for the interference induced by supply-constrained resource allocation in a marketplace \citep{munro2021treatment, wager2021experimenting}. In both cases, we see that understanding the underlying resource allocation dynamics can be beneficial in designing more efficient statistical procedures. 


\section{The System Model}
\label{sec:model}

We now formally introduce the system model, illustrated in Figure \ref{fig:sysdiag}. 

{\bf System Primitives.} We consider a service system operating in continuous time $t \in \mathbb{R}_+$, consisting of $K$ parallel servers labeled $\{1,\ldots,K\}:=\{1, \ldots, K\}$. Jobs arrive to the dispatcher according to a Poisson process of rate $\lambda$. An incoming job first arrives at a dispatcher, who subsequently immediately directs the job to queue for service in one of the $K$ servers according to a control policy. We denote by $Q_i(t)$ the queue length at server $i$ at time $t$, defined by the total number of jobs waiting including any currently in processing, and by $Q(t)=(Q_1(t),\ldots,Q_K(t))$ the queue lengths vector. 

We assume that each job is associated with a size that is independently and identically distributed according to an exponential distribution with rate $1$. Under normal operation, a server $i$ performs work at a rate $\mu_i=1$, and a job departs from the system once it has received the amount of work equal to its size. As such, the time a job at server $i$ would spend in service is a random variable distributed according to  $\mbox{Exp}(1)$. We assume that servers process jobs in a first-come-first-serve (FIFO) order, with no preemption. We denote by $\rho$ the system load, defined to be the ratio between the arrival rate and total service capacity:  $\rho=\frac{\lambda}{\sum_{i} \mu_i}$.

We assume that when a server experiences slowdown, its service rate drops to a reduced level of $\alpha < 1$ for a slowed down server, with $\alpha \in (0,1)$ being the {slowdown factor}. We assume that the system remains under-loaded, with $\rho<1$ after a possible server slowdown. Specifically:
\begin{equation*}
    \rho=\twopartdefnoif{\frac{\lambda}{K}}{\mbox{before slowdown}}{\frac{\lambda}{K-1+\alpha}}{\mbox{after slowdown}},
\end{equation*}
i.e., we require that 
\begin{equation*}
    \lambda<K-1+\alpha.
\end{equation*}
The rationale is that, should this not be the case, the queue lengths at some of the servers would be so large that the presence of a slowdown would be all but obvious to the operator. Therefore, as a first step, we will focus on this stable setting where both the presence of a slowdown \emph{and} the identity of the slow server is difficult to detect. 

Finally, we assume that up to one server may experience slowdown, and such an event would have occurred prior to $t=0$, so that the system would have reached the new, post-slow-down equilibrium by $t=0$. We believe this to be a reasonable assumption in systems where the slowdown events are relatively rare so that any temporal carryover effects associated with shifting between two system equilibria are of second order in the ensuing statistical analysis. Removing this stationarity assumption so that a slowdown could occur in the course of the data is an interesting variation, which we will defer to future research. 

{\bf Adaptive Congestion Control Policy.} A {congestion control policy}, or simply control policy, refers to the rule used by the dispatcher to assign incoming jobs to servers. We will focus on adaptive control policies that use the queue length vector $Q(t)$ as their main input when making assignment decisions. We will also allow the policy to  access  some form of randomization, where the decision at time $t$ can depend on some random variable $\xi(t)$, which is i.i.d.~for all $t$. The control policy is formalized by a map 
\begin{equation}
	R:\mathbb{Z}_+^K \times \mathbb{R}\rightarrow \{1, \ldots, K\},
	\label{eq:controlpol}
\end{equation}
 such that if a job arrives at time $t$, it is sent to the sever with index $R(Q(t^-),\xi(t)).$  One representative control policy in this class is the {join-the-shortest-queue (JSQ)} policy, whereby the dispatcher sends the job to the server with a shortest queue, with ties broken according to some predetermined rule (e.g., random tie-breaking). The {random assignment} policy is another popular choice which trivially belongs to this class: each job is routed to a server sampled uniformly at random, regardless of the queue lengths. 
 
 Note that, with a control policy conforming to \eqref{eq:controlpol}, the queue length process $\{Q(t)\}_{t\in \mathbb{R}_+}$ forms  a homogeneous, irreducible, continuous-time Markov Chain on the countable state space $\mathcal{S}=\mathbb{Z}^K_+$.

{\bf Data Collection.} The decision maker has access to observational congestion data collected in the format of an $K\times N$ panel denoted by $\calQ$, illustrated in Figure \ref{fig:datadiag}. The panel consists of $N$ samples of queue lengths collected at equally spaced intervals, $t_1 < t_2 < \ldots < t_N$: $\calQ = (Q(t_1), Q(t_2), \ldots, Q(t_N))$, where $t_1 =0$ and $t_i = L(i-1)$ for $i = 2, \ldots, N$.  The constant $L >0$ is referred to as the sampling interval. We will denote by  $Q_{k,n}$ the queue length at server $k$ in the $n$th sample in $\calQ$.

{\bf Summary Statistic and Decision Rule.} We consider a decision maker who uses a set of summary statistics that is intended to serve as an intuitive summary of the raw observational data $\calQ$ and guide subsequent detection decisions. 
\begin{definition}[Summary Statistic] 
\label{def:sum_stat}
A summary statistic is defined as a function $s: \mathbb{R}_+^{K \times N} \to \left(\mathbb{R}_+\cup\{\infty\} \right)^K$, which maps the observational data to a $K$-dimensional vector. 
\end{definition}
We can think of $s_k(\cdot)$ as a summary score that the statistic associates with the $k$th server. For instance, the empirical mean of the congestion data for this server would be a natural example of such a statistic.  

Given the values of the summary statistic computed on the observational data, $s(\calQ)$, the decision maker then employs a decision rule to arrive at a final output. While it is in theory possible to simply combine the summary statistic and decision rule into a single mapping, defining them separately allows for the possibility of using the same decision rule over different summary statistics, and vice versa, as we will  in later parts of the paper. We will denote by $\calD$ the set of all possible  final outputs, which we call the decision space.

\begin{definition}[Decision Rule]
\label{def:dec-rule}
A decision rule, $r:  \mathbb{R}^K \to \calD$ takes as input a $K$-dimensional summary statistic and outputs an element in the decision space, $\calD$. 
\end{definition}
Below is an example of a natural decision space as well as a class of decision-rules that will be useful for our analysis. 
\begin{definition}[Index-Plus Decision Space] 
	\label{def:index-plus}
	The index-plus decision space is defined to be the set $\{1, \cdots, K\} \cup \{\emptyset\}$, consisting of the index of all $K$ servers plus the null symbol $\emptyset$. 
\end{definition}
While using an index-plus decision space, an output within $\{1, \cdots, K\}$ corresponds to the decision rule declaring the identity of the slowdown server, whereas the output $\emptyset$ corresponds to the decision rule declaring the analysis to be inconclusive.

\begin{definition}[Relative Threshold Rule]
	\label{def:rel_threshold}
	Fix $\gamma >0$ and an index-plus decision space. A majorizing relative threshold rule is defined by: 
	\begin{enumerate}
            \item $r(s) = k$, if there exists $k \in \{1, 2, \ldots, K\}$ such that 
		\begin{equation}
		   s_k > (1+\gamma) s_j, \quad \forall j \neq k     
     \label{eq:major_rel_threshold}
		\end{equation}
		\item $r(s) = \emptyset$, otherwise,
	\end{enumerate} 
Similarly, fixing $\gamma \in (0,1)$, a minorizing relative threshold rule is defined by changing  condition in \eqref{eq:major_rel_threshold} to: 
		\begin{equation*}
		   s_k < (1-\gamma) s_j, \quad \forall j \neq k. 
		\end{equation*}
\end{definition}	

 That is, an index $k$ is selected in a relative threshold rule if its statistic dominates (or is dominated by) all other indices by a sufficient multiplicative margin. The rationale behind the relative threshold rule is that in practice, acting upon a server slowdown often means triggering a process that involves costly diagnosis and engineering hours. If a wrong server is identified, these resources would be wasted. Therefore, declaring a server slowdown should be contingent upon the data showing a substantial, proportional difference in the summary statistics across the servers. 

\begin{remark}
\label{rem:majorminor_thresh} Note that the sets of all majorizing and minorizing threshold rules, respectively, are mathematically equivalent to each other, in that a majorizing rule using $s_k$ and $\gamma$ is equivalent to a minorizing rule using $s'_k := 1/s_k$ (with $s'_k=\infty$ if $s_k=0$) and $\gamma' := 1-1/(1+\gamma)$, and vice versa. Making a distinction between the two sets of rules, however, is more convenient as one version is often more intuitive than another in specific instances. 
\end{remark}



Finally, to facilitate our exposition, we will focus on the \emph{symmetric} mappings, including the congestion control policy, summary statistics and decision rules. A mapping is symmetric if it does not depend on the identity of the servers beyond their operating statistics as measured in $\calQ$. For instance, control policies such as JSQ and Replicate-to-the-Shortest-Queues (RSQ) \cite{atar2019replicate} are naturally symmetric when they are coupled with a randomized-tie breaking rule, whereas a version of JSQ  where ties are broken by favoring servers with a lower index $k$ would not be.

\section{Failures of Marginal Congestion Statistic: Independent Samples Analysis}
\label{sec:queue_lengths}

We show in this section how and why the presence of adaptive control policies  could render marginal congestion statistics ineffective for slowdown detection, especially when the system is heavily loaded (post slow down). The heavy traffic regime is practically relevant because in this case continuing to operate with a slow server is likely to induce significant congestion. We will relax the heavy traffic assumption in Section \ref{sec:finte_time_pa} when analyzing the finite-time performance.  


We will begin by defining the following problem setting where there is at most one slow server and that the decision maker knows the magnitude of a potential slowdown, $\alpha$. We consider the case of an unknown $\alpha$ numerically in Section \ref{subsec:sim_reliability_unknown}.

\begin{definition}[One-Server Failure with Known Slowdown Rate] 
\label{def:known_slowdown}
Fix $K$ and a slowdown factor $\alpha <1$. The true state of nature can be one of the following two scenarios (hypotheses): 
\begin{enumerate}
    \item C1: Server 1 slows down and operates at a rate of $\alpha$, while all other $K-1$  servers operate at unit rate. 
    \item C2: There is no slowdown and all $K$ servers operate at unit rate. 
\end{enumerate} 
The decision space is index-plus as in Definition \ref{def:index-plus}. We say that there is a \emph{success} if and only if: 
\begin{enumerate}
    \item Under C1: decision rule correctly identifies the  slow server,   with $r(s(\calQ)) = 1$. 
    \item Under C2: decision rule declares the analysis to be inconclusive (i.e., do nothing) , with $r(s(\calQ)) = \emptyset$. 
\end{enumerate} 
\end{definition}
We now define the key metric of {reliability}. 

\begin{definition}[Reliability]
    The reliability is the minimum probability of success across all states of the world. In the case of the setting of Definition \ref{def:known_slowdown}, it amounts to: 
\begin{equation*}
  p_+ := \min_{C \in \{C1, C2\} }  \pb_C(\mbox{success}). 
\end{equation*}
\end{definition}

\begin{remark}
\label{rem:majorminor_thresh} 
Notice that in Definition \ref{def:known_slowdown} we only allow server 1 to be the slower server. This is in fact without loss of generality since we are largely concerned with symmetric mappings (Section \ref{sec:model}). The symmetry helps us simplify analysis by focusing on a binary hypothesis setting. However, when policies are asymmetric, it should be relatively straightforward to extend Definition \ref{def:known_slowdown} by considering, in place of C1, $K$ distinct hypothesis, where each corresponds to one of the $K$ server slows down while the others remain healthy. 
\end{remark}

For data collection, we will consider a simplified setting where the $N$ queue length vector samples, $Q_{\dot, n}$, are drawn i.i.d.~from the steady-state distribution of $\{Q(t)\}_{t\in \mathbb{R}_+}$. This can be thought of a limiting regime as the sampling interval $L$ tends to infinity, so that each consecutive samples become asymptotically independent from one another. Not only does the i.i.d.~analysis help to sharpen the underlying insight, considering that correlations between samples generally increase the variance of the underlying estimator (cf.~\cite{jones2004markov}), showing how marginal statistics can fail even when operating under i.i.d.~data would further strengthen the point about their fragility. We will consider the more general setting of correlated samples (finite $L$) in Section \ref{sec:finte_time_pa} for the performance of the proposed potential action statistics.

We will consider the family of admissible marginal summary statistics, defined as follows. 

\begin{definition}[Marginal Summary Statistic]
\label{def:marginal_stat}
A marginal summary statistic is one in which the $k$th coordinate of the summary statistic (Definition \ref{def:sum_stat}) is a function only of the data collected from the $k$th server. Specifically,  $s$ is marginal summary statistic, if there exists a function $g:\mathbb{R}_+^N \rightarrow \mathbb{R}$, such that 
\begin{equation*}
	s(\calQ) = (g(Q_{1, \cdot} ), g(Q_{2, \cdot}), \ldots, g(Q_{K, \cdot}) ), \quad \forall \calQ.  
\end{equation*}
We will use $g$ to denote the corresponding marginal summary statistic.  A marginal statistic is \emph{admissible} if the function $g$ is continuous and scale-invariant in the sense that  for all $c\geq 0$, $g(cx)=cg(x)$ for all $x \in \mathbb{R}_+^N$.
	We will denote by $\mathcal{G}$ be the set of all  admissible marginal statistics. 
 \end{definition} 
 
 Admissible marginal statistics encompass many simple and intuitive summary statistics. Examples include the arithmetic and geometric means of the empirical distribution associated with $Q_{k, \cdot}$, as well as the $L^p$ norms of $Q_{k, \cdot}$.

The following theorem is the main result of this section. It shows that under \emph{any} marginal statistics the reliability of learning slowdown is always vanishingly small in the heavy traffic regime. While we state the theorem under the JSQ policy, we in fact prove a more general statement that applies to a broader set of congestion control rules that induce state-space collapse in the queue length distribution; the JSQ policy is one such example.  

The main insight behind the proof is that under heavy traffic, the adaptive routing produced by  JSQ are so effective that the queue lengths across the different servers are highly similar, despite the variation in the underlying service rates. This subsequently causes any marginal statistics to perform poorly. The proof is given in Appendix \ref{app:thm:no_ql_detection}.  

\begin{theorem}
\label{thm:no_ql_detection}
Consider the problem setting in Definition \ref{def:known_slowdown}. Fix any admissible marginal statistic $g\in\mathcal{G}$ and a relative threshold decision rule satisfying Definition \eqref{def:rel_threshold}. Suppose the congestion control policy is JSQ. Then for {any} $\alpha \in (0,1)$ and $N \in \mathbb{N}$, the reliability vanishes in heavy traffic: 
\begin{equation*}
	p_+  = \min_{C \in \{C1, C2\} } \pb_C(\mbox{success}) \to 0, \quad \mbox{as } \lambda \to \alpha + (K-1). 
\end{equation*}
\end{theorem}


\section{Potential Action Statistic: Independent Samples Analysis}
\label{sec:action}

We have seen in the previous section how marginal statistics may fail even under i.i.d.~data samples. We now show how under the same setting as in Theorem \ref{thm:no_ql_detection}, by using a new summary statistics called potential actions, we are now able to achieve vastly improved slowdown detection.  The analysis in this section will continue to use the i.i.d.~sample assumption ($L=\infty$) because it helps clarify the key insights. Next, in Section \ref{sec:finte_time_pa}, we show how to remove the i.i.d.~assumption and extend the performance guarantees of the potential action statistics to the general setting with finite sampling interval $L<\infty$.

To motivate the new summary statistics, recall from the previous subsection that marginal statistics fail because the adaptive routing actions by JSQ tend to equalize the queue lengths and thus obfuscate the underlying service rate variations. The new statistic addresses this by instead focusing on capturing the routing behaviors themselves. The key insight here is that any performant congestion control policies that can successfully stabilize the system under a heavy traffic load must, by definition, automatically adapt their routing actions to match the underlying service rates. It therefore suggests that tracking the changes in routing actions should provide us with a much more informative statistic. 

There is one more challenge: the dataset that we have access to consists only of observed queue lengths and not of routing actions. To solve that, we will instead construct, using only the queue lengths, a ``potential'' action vector that represents the routing probabilities for a hypothetical arrival, as follows: 

\begin{definition}[Potential Action] 
\label{def:po_action}
Let $R$ be the routing map and $\xi$ the random seed in \eqref{eq:controlpol}. A potential action $\hat R \in \rp^K$ associated with a vector  $q \in \rp^K$ is the routing probabilities for a hypothetical job arriving to the system with queue lengths $q$. In particular,   
\begin{equation}
    \hat R(q) = \left(\pb(R(q, \xi) = k)\right)_{k = 1, \ldots, K}. 
    \label{eq:PA_def}
\end{equation}
\end{definition}

\begin{definition}[Potential Action Summary Statistic] 
	\label{def:count_pa}  The potential action summary statistic, $h$, is defined  as: 
	\begin{equation*}
		h_k(\calQ) = \frac{1}{N} \sum_{n=1}^N \left[\hat R(Q_{\cdot,n})\right]_k, \quad k=1, \ldots, K. 
	\end{equation*}
	\end{definition}

\subsection{Successful Detection Using Potential Action Statistics}
\label{subsec:PA_known_alpha}

We will next state our main result for this section showing that slowdown can be successfully detected under the i.i.d.~data setting using potential actions. The result will apply to a family of congestion control policies that contains the JSQ as a special case, defined as follows: 
\begin{definition}[Maximally Stable Control Policy]
	\label{def:maxStable}
	A congestion control policy is maximally stable if whenever $\lambda < \sum_i \mu_i$, the system remains stable: the Markov chain  $\{Q(t)\}_{t\in \mathbb{R}_+}$ is positive recurrent, and admits a unique steady-state distribution. 
\end{definition}
In addition to the JSQ, the Power-of-Memory \citep{shah2002use} and Persistent-Idle \citep{atar2020persistent} policies are known to be maximally stable.\footnote{Strictly speaking, these two require the use of a small, finite memory to store summaries of past dispatch decisions, in addition to the current queue length. However, all of our results can be easily extended to the setting where the control policy uses a small finite memory.}, while Join-the-Idle-Queue \citep{lu2011join} and Power-of-Choice \citep{vvedenskaya1996queueing,mitzenmacher2001power} policies, on the other hand, are known not to be \citep{atar2020persistent}. 

With a slight abuse of notation, define $ \hat R_{n}$ to be the potential action associated with the $n$th queue length sample, and $\pi^R$ to be its steady-state mean: 
\begin{equation*}
    \hat R_{n} = \hat R(Q_{\cdot, n}), \quad \pi^R = \EE\left[\hat R_1 \right]. 
\end{equation*}
Note that $\EE\left[h_k(\calQ)\right] = \pi^R_k$ as a result of the stationarity assumption. 

Consider the problem setting of Definition \ref{def:known_slowdown} where server 1 may slowdown. The next result is a key lemma that characterizes $\pi^R$ under hypothesis C1. The proof is given in Appendix \ref{app:lemma:ub and lb on ratio}. 
\begin{lemma}
	\label{lemma:ub and lb on ratio}
   Fix $\alpha \in (0,1]$. Suppose C1 is true, and $\alpha <\lambda<\alpha+K-1$. Then under any maximally stable and symmetric control policy, the following holds. 
   \begin{enumerate}
       \item Let $\lambda_k$ be the stationary arrival rate of jobs to server $k$. Then
       \begin{equation}
            \pi^R_k = \lambda_k/\lambda, \quad k = 1, \ldots, K. 
            \label{eq:PASTA_PA}
        \end{equation}
        \item $\pi^R_k=\pi^R_2$ for all $k>1$ and 
\begin{equation}
    \alpha\left(1 -(1-\rho)\frac{K-1+\alpha}{\alpha}\right) \leq \frac{\pi^R_1}{\pi^R_2} \leq \alpha \left ( \frac{1}{\rho-(1-\rho)\frac{\alpha}{K-1}}\right). 
    \label{eq:piR_upperlower}
\end{equation}
\item As $\rho \to 1$ (i.e., $\lambda \to \alpha+K-1$), we have
\begin{equation*}
      \pi^R_1 \to  \frac{\alpha}{K-1+\alpha},  \quad    \pi^R_2 \to \frac{1}{K-1+\alpha}, \quad \frac{\pi^R_1}{\pi^R_2} \to \alpha. 
\end{equation*}
   \end{enumerate}

\end{lemma}

Lemma \ref{lemma:ub and lb on ratio} shows that under a maximally stable congestion control policy, the identity of the slower server is meaningfully reflected in the mean potential action, as we had conjectured earlier. In fact, as the load becomes heavier, the ratio between the mean potential actions between a slower and healthy server converges to the slowdown factor $\alpha$. 

Moreover, Lemma \ref{lemma:ub and lb on ratio} also provides clear guidance on how to design an effective relative threshold rule (Definition \ref{def:rel_threshold}) using potential action statistics. Note that the ratio $\frac{\pi^R_1}{\pi^R_k}$ for $k>1$ is 1 when no servers slowdown, but drops to approximately $\alpha<1$ when server 1 does slow down. Then, in order for the minorizing relative threshold rule to be able to differentiate these two cases, a reasonable choice would be to set the threshold $(1-\gamma)$ to be the mid-point between $\frac{1+\alpha}{2}$, which yields the choice $\gamma = 1-\frac{1+\alpha}{2} = \frac{1-\alpha}{2}$. 

The next theorem, the main result of this section,  confirms that this is indeed effective. The proof is given in Appendix \ref{app:thm:pa_detection_simple}. 
\begin{theorem}
	\label{thm:pa_detection_simple}
 Consider the problem setting in Definition \ref{def:known_slowdown}.  Suppose the congestion control policy is symmetric and maximally stable. Fix the summary statistics to be that of potential action and the decision rule a minorizing relative threshold rule with 
 \begin{equation*}
     \gamma=\frac{1-\alpha}{2}.
 \end{equation*} 
 Then, for all $N >0$ 
	\begin{equation*}
		      p_+ \geq 1- \max \left\{ (K-1) \exp\p{-\frac{(1-\alpha)^2}{32K^2}N}, \, K\exp\p{-\frac{(1-\alpha)^2}{8K^2}N}  \right\}. \nonumber,  
	\end{equation*}
 as  $\lambda \to \alpha + K-1$. 
\end{theorem}

Theorem \ref{thm:pa_detection_simple} stands in sharp contrast to Theorem \ref{thm:no_ql_detection}. We see that the potential action statistic in combination of a relative threshold rule is  able to succeed with overwhelming probability as the same size grows, whereas under the same heavy-traffic regime, using any  marginal statistic and  relative threshold decision rule would lead to vanishingly small chance of success.

\section{Finite Time Analysis of Potential Action}
\label{sec:finte_time_pa}

We now extend the analysis of the potential action statistic and consider the original setting where the sampling interval $L$ is finite. In this case, the $N$ samples collected are not independent from one another, thus making the statistical analysis much more challenging. The main goal for this section is to establish a central limit theorem (CLT) for the potential action statistic, and further use the CLT to approximate the reliability of potential actions in a finite time setting. The key takeaway is that the effectiveness of the potential action statistics extends to the finite-time, non-i.i.d.~setting as well.

Let us briefly recall the setup for the data collection process in Section \ref{sec:model}. We assume that the load balancer uses the join-the-shortest-queue (JSQ) policy with ties broken uniformly at random, and that the system is under-loaded, so that $\rho = \lambda/ \sum_{i} \mu_i < 1$. As discussed in Section \ref{sec:model}, the system operating under JSQ will be stable whenever $\rho<1$, so that the queue length process $Q(t)$ corresponds to a positive-recurrent Markov chain. We assume the queue length process is initialized in its stationary distribution at $t=0$. The potential action is $\hat R_n = \hat R(Q((n-1)L))$, where $L \in (0, \infty)$ is the sampling interval.  

Because the underlying queue length process takes values in an countably infinite state space, conventional Markov chain CLTs do not apply. The technical contribution the following theorem is to establish that the queue length dynamics are geometrically ergodic under the JSQ policy when service rates are non-homogeneous, where the novelty is the construction and analysis of an appropriately chosen Lyapunov function. This result will then allow us to obtain the main CLT for the potential action statistics, using the tools developed in \citep{meyn1993stability, ibragimov1962some, jones2004markov, gallegos2024equivalences}.  We have the following theorem; the proof is given in Appendix \ref{app:thm:clt_pa_finite_time}. 

\begin{theorem}[Central Limit Theorem for linear combinations of the Potential Action Statistic]
\label{thm:clt_pa_finite_time}
    Fix $\lambda$ and $\mu_1,\ldots,\mu_K$ such that $\rho \in (0,1)$ and $\min_i \mu_i>0$. Fix ${\eta}=(\eta_1,\ldots,\eta_K)\in \mathbb{R}^K$ and let $h(\calQ)=(h_1(\calQ),\ldots,h_K(\calQ))$. Suppose the congestion control policy is JSQ. Then,  
    \begin{equation*}
        \sqrt{N}\p{\eta^T h(\calQ) -\eta^T  \pi^R } \Rightarrow \mathcal{N}(0, \sigma^2_{{\eta}}),
    \end{equation*}
    as $N\to \infty$, where $\sigma_{{\eta}}^2  :=  \var(\eta^T \hat R_{1})  + 2\sum_{n=2}^\infty \cov(\eta^T  \hat R_{1}, \eta^T \hat R_{n} ) < \infty$. 
\end{theorem}

We use Theorem \ref{thm:clt_pa_finite_time} to approximate probabilities related to the success or failure of the relative threshold test under different hypotheses, where the vector $\eta$ is chosen depending on the context.
\begin{remark}
    For relatively complex Markov chains, such as ours, it is well known that it is often difficult to derive an explicit formula for the variance $\sigma_{\beta}^2$ due to the infinite sum of covariances. In fact, existing quantitative bounds on the rate of the geometric convergence of the Markov chain to the stationary distribution are often too loose to be of practical use; see, for instance, discussions in \citep{meyn1994computable, rosenthal2002quantitative}. As a result,   there is a large body of literature devoted to estimating $\sigma_{\beta}^2$  via numerical means such as Markov Chain Monte Carlo simulation \citep{geyer1992practical, glynn1992asymptotic, robert1995convergence}. The simulation-based approach is likely feasible for a practitioner in our setting as well, who seeks to use an estimate of $\sigma_{\beta}$ in order to choose a sample size $N$ that satisfies a desired level of reliability (see Theorem \ref{thm:normal_appox_reliability}). 
\end{remark}

\begin{remark}
    Note that Theorem \ref{thm:clt_pa_finite_time} holds for any congestion control policy under which the underlying Markov Chain is geometrically ergodic, not just JSQ.
\end{remark}

\subsection{Reliability of Potential Action using Normal Approximation}

Given a vector $\eta \in \mathbb{R}^K$, The CLT in Theorem \ref{thm:clt_pa_finite_time} provides the justification for the normal approximation
\begin{equation}
    \sqrt{N} (\eta^T h(\calQ) - \eta^T \pi^R)  \approx \calN(0 , \sigma^2_{\eta})
    \label{eq:normal_approx}
\end{equation}
for large $N$.  We will use the above normal approximation to to derive non-asymptotic reliability guarantees for a relative threshold rule using potential actions, similar to the one analyzed in Theorem \ref{thm:pa_detection_simple}, but now without the i.i.d.~sampling nor heavy-traffic asymptotics.  We will consider the problem setting of a single server slowdown with known $\alpha$ (Definition \ref{def:known_slowdown}). We will choose a specific vector $\eta$ corresponding to the events whose probabilities we wish to approximate.  

Consider a minorizing threshold rule with $\gamma \in (0,1)$ (Definition \ref{def:rel_threshold}). Recall that a key ingredient for this decision rule is the event $\{h_k(\calQ) < (1-\gamma) h_j(\calQ)\}$, which we can rewrite as: 
\begin{align}
     & h_k(\calQ) <  ( 1-\gamma) h_j (\calQ) \nln
\Leftrightarrow  & \sqrt{N}(h_k(\calQ) - \pi^R_k) <  \sqrt{N}\left[ (1-\gamma) (h_j(\calQ) - \pi^R_j + \pi^R_j) - \pi^R_k\right] \nln
\Leftrightarrow  &  \sqrt{N}(h_k(\calQ) - \pi^R_k) - (1-\gamma)  \sqrt{N}(h_j(\calQ)  - \pi^R_j) <  \sqrt{N}\left[  (1-\gamma) \pi^R_j - \pi^R_k\right] 
\label{eq:h_k_event}
\end{align}
Using \eqref{eq:normal_approx}, we approximate the left hand side by a normal random variable $\bar{H}_{kj}$ as follows:
\begin{equation}
    \sqrt{N}(h_k(\calQ) - \pi^R_k) - (1-\gamma)  \sqrt{N}(h_j(\calQ)  - \pi^R_j) \approx  \bar{H}_{kj} \overset{d}{=} \calN(0 , \sigma^2_{k \to j}),
    \label{eq:specific_kj_approx}
\end{equation}
where $\sigma^2_{k \to j}$ is defined to be the variance given in Theorem \ref{thm:clt_pa_finite_time} for the specific choice of $\eta$ such that:
\begin{equation*}
    \eta_i=\threepartdefelse{1}{i=k}{-(1-\gamma)}{i=j}{0}
\end{equation*}
Thus, using \eqref{eq:h_k_event} and \eqref{eq:specific_kj_approx}, the probability of the event $\{h_k(\calQ) < (1-\gamma) h_j(\calQ)\}$ is approximated by that of
\begin{equation}
    \calE_{kj} := \{\bar{H}_{kj}   <  \sqrt{N}\left[  (1-\gamma) \pi^R_j - \pi^R_k\right] \},
    \label{eq:HkH_j_event}
\end{equation}
where the variable  $\bar{H}_{kj}$ is normally distributed with mean $0$, and variance $\sigma^2_{k \to j}$. Therefore:
\begin{equation*}
    \mathbb{P}(\calE_{kj})= \Phi \left( \sqrt{N}\left[  (1-\gamma) \pi^R_j - \pi^R_k\right]/\sigma^2_{k \to j}\right),
\end{equation*}
where $\Phi( \cdot)$ is the cdf of a standard normal random variable.

 Note that the event $ \cap_{j\neq k}\calE_{k j}$   implies that the minorizing relative threshold rule would output $k$ as the slow server,  whereas the event $\cap_k \cup_{j\neq k} \overline \calE_{kj}$ implies the decision rule would output $\emptyset$. 


The above derivation leads to the following main result of this subsection, a normal approximation for the performance of a relative threshold test. We will denote by $\bar p_+$ the reliability of a procedure under the normal approximation of \eqref{eq:normal_approx}, recalling that the hypotheses C1 and C2 correspond to the slowdown of server 1 or no slowdown at all, respectively. The proof is given in Appendix \ref{app:thm:normal_appox_reliability}. 

\begin{theorem}
\label{thm:normal_appox_reliability}
Consider the problem setting in Definition \ref{def:known_slowdown}. Suppose the congestion control policy is JSQ with random tie-breaking. Fix the summary statistics to be that of potential action and the decision rule a minorizing relative threshold rule with 
\begin{equation*}
     \gamma = \frac{1-1.1\alpha}{2}. 
\end{equation*}
Then, for all $L >0$, $K \geq 3$, $\alpha < 0.9$ and   $\lambda \geq 0.95 (\alpha + K-1)$, the reliability under  normal approximation satisfies: 
\begin{align*}
    \bar p_+ = & \min_{C \in \{C1, C2\}} \pb_{C}(\mbox{success}) \nln
&\geq  1- \max\left\{  (K-1)\Phi\left( -  \sqrt{N} \frac{1-1.1\alpha}{2K  \sigma_{C1, 1 \to 2}}   \right) 
, K  \Phi\p{ -\sqrt{N}\frac{1-1.1\alpha}{2K \sigma_{C2,1\to 2}}   } \right \}, 
\end{align*}
where $\sigma_{C1, k\to j}$ and $\sigma_{C2, k\to j}$ represent the values of $\sigma_{k \to j}$ under hypotheses $C1$ and $C2$, respectively, in Definition \ref{def:known_slowdown}. 
\end{theorem}
The theorem shows that the reliability converges exponentially fast to 1 as $N\to \infty$, as we would want for an effective decision rule. The non-asymptotic nature of the bound also provide a practical way for finding the optimal sample size $N$ for a desirable level of confidence. Finally, comparing the above characterization with that in Theorem \ref{thm:pa_detection_simple}, we see that not only have we removed the assumption of i.i.d.~samples, but we also obtained a non-asymptotic characterization that does not require the heavy traffic limit (i.e., $\lambda \to \alpha + K-1$).


\section{Numerical Results}\label{sec:simulations}
The purpose of this section is to provide numerical results that support the use of potential actions as a summary statistic for the purpose of service slowdown detection.

In Section \ref{subsec:sim_theta_vs_alpha} we complement the result of Lemma \ref{lemma:ub and lb on ratio} by estimating the dependence of the potential action probabilities as a function of the load $\rho$ and the slowdown factor $\alpha$. We show that the ratio between the routing probability to the slow server and other servers is even closer to the slowdown factor $\alpha$ than our bounds indicate, suggesting that potential actions may provide a strong signal for detection.
In Section \ref{subsec:sim_reliability_known} we complement the results of Theorems \ref{thm:pa_detection_simple} and \ref{thm:normal_appox_reliability} by estimating the reliability of the relative threshold test based on potential actions as a function of $\alpha$, $L$ and $N$. We show that the sampling interval $L$ has a profound impact on reliability for small values of $L$ such that a larger interval would require a substantially smaller number of samples to achieve the same reliability.
In Section \ref{subsec:sim_reliability_unknown} we demonstrate that our approach can be beneficial in situations where $\alpha$ is unknown apriori. In particular, we demonstrate that choosing a minimal slowdown magnitude and designing the test accordingly results in good performance for more substantial slowdowns.
Finally, in Section \ref{subsec:sim_example_use} we show how a practitioner can use potential actions for slowdown detection as demonstrated in Figure \ref{fig:dashboard_intro} in the introduction.

\subsection{Setup}\label{subsec:sim_setup}
For all the simulations in this section we simulate a discrete time parallel server system with $K$ servers, each with its own buffer in which a queue can form, and one dispatcher. The service rate of server $i$ is given by $0<\mu_i<1$, and the arrival rate is given by $0<\lambda<1$. Thus, the inter-arrival and service times are geometrically distributed. The load is defined as $\rho=\lambda/\sum \mu_i$. At each round, the order of events is as follows:
\begin{enumerate} 
    \item With probability $0<\lambda<1$ a job arrives to the dispatcher and is sent to the server with the least number of jobs to process (i.e. JSQ). In all simulations, ties are broken randomly.
    \item If server $i$ has jobs to process, then with probability $\mu_i$ it completes one job.
\end{enumerate}

If a change in server speed occurs at some round, then the corresponding service rate $\mu_i$ changes to $\alpha \mu_i$.

\subsection{Potential Routing Probabilities and the Slowdown Factor}\label{subsec:sim_theta_vs_alpha}

In Section \ref{sec:action}, Lemma \ref{lemma:ub and lb on ratio}, we argued that the ratio between the routing probabilities $\pi^R_1$ and $\pi^R_2$, which can be estimated effectively using potential actions, is also close to the slowdown factor $\alpha$ for high enough loads. This enabled us to use potential action not only to detect that a change occurred and in which server, but also derive information regarding the magnitude of the slowdown.

We now demonstrate that the ratio $\pi^R_1/\pi^R_2$ is even closer to the slowdown factor $\alpha$ 
than what the theoretical bounds in Lemma \ref{lemma:ub and lb on ratio} indicate. The implication is that for moderate to high loads, cases where quick detection matters most, potential actions provide a powerful signal for inference.

\begin{figure*}[t!]
    \centering
    \includegraphics[width=0.7\textwidth]{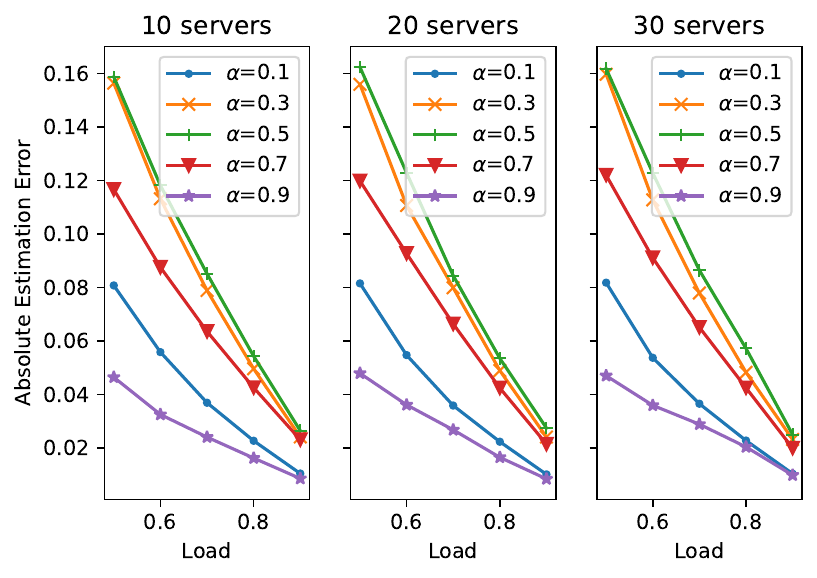}
    \caption{The absolute error between the estimated $\pi^R_1/\pi^R_2$ ratio and the actual slowdown factor $\alpha$, for different number of servers and loads. The error is relatively small for moderate to high loads in all cases, suggesting that potential action based statistics can be beneficial not only to detect slowdowns, but also provide information on the slowdown magnitude. }
    \label{fig:1_theta_vs_alpha}
\end{figure*}

We choose values for $K$, $\alpha$ and $\rho$, such that $K\in\{10,20,30\}$, $\alpha\in \{0.1, 0.3, 0.5, 0.7, 0.9\}$ and $\rho \in \{0.5, 0.6, 0.7, 0.8, 0.9\}$. For each combination of $K$, $\alpha$ and $\rho$ we run a simulation of the parallel server system to obtain a long-run estimation of $\pi^R_1/\pi^R_2$ and then compare it to the value of $\alpha$. The results are depicted in Figure \ref{fig:1_theta_vs_alpha}. 

For each value of $K$, we plot the absolute error between the estimated $\pi^R_1/\pi^R_2$ value and the value of $\alpha$, as a function of the load $\rho$. The plots are similar for the different number of servers and we can see that the error does not exceed 0.165 and approaches zero as the load increases. In addition, we find that the error is not monotone in $\alpha$, were $\alpha=0.5$ has the largest errors, and $\alpha\in\{0.1,0.9\}$ has the smallest ones. For moderate to high loads, we see that the error corresponds to just a few percentages of slowdown, e.g., for a load of $0.8$, and a slowdown of $50\%$ ($\alpha=0.5$), the estimated ratio lies in $50\% \pm 6\%$.

\subsection{Reliability for Known Slowdown Factor}
\label{subsec:sim_reliability_known}
The purpose of this section is to complement the results of Theorems 
\ref{thm:pa_detection_simple} and \ref{thm:normal_appox_reliability} by demonstrating how the reliability of the relative threshold test based on potential actions behaves as a function of the slowdown factor $\alpha$, the number of samples 
$N$ and the time between samples $L$. We emphasize that the test threshold is chosen based on the \textit{known} slowdown factor $\alpha$. Section \ref{subsec:sim_reliability_unknown} demonstrates how the reliability behaves when the threshold is chosen for a specific $\alpha$ but the actual slowdown factor is different. 

We fix the number of servers $K=10$ and consider $\alpha \in \{0.5,0.7,0.85\}$. For each $\alpha$ we set the arrival rate to be $\lambda=0.95K/(K-1+\alpha)$ which corresponds to a load of $\rho=0.95$ under the hypothesis that a server slowed down. To evaluate Reliability, we consider two cases: (1) no slowdown, in which case $\mu_i=1/K$ for all $i$, and (2) server 1 slowed down, in which case $\mu_1=\alpha/K$ and $\mu_i=1/K$ for all $i\geq 2$.
We consider different values for $N$ and $L$ and use a Monte Carlo simulation to approximate the reliability under the relative threshold test using potential actions and $\gamma=(1-1.1\alpha)/2$ as in Theorem \ref{thm:normal_appox_reliability}.

\begin{figure*}[h!]
    \centering
    \includegraphics[width=1\textwidth]{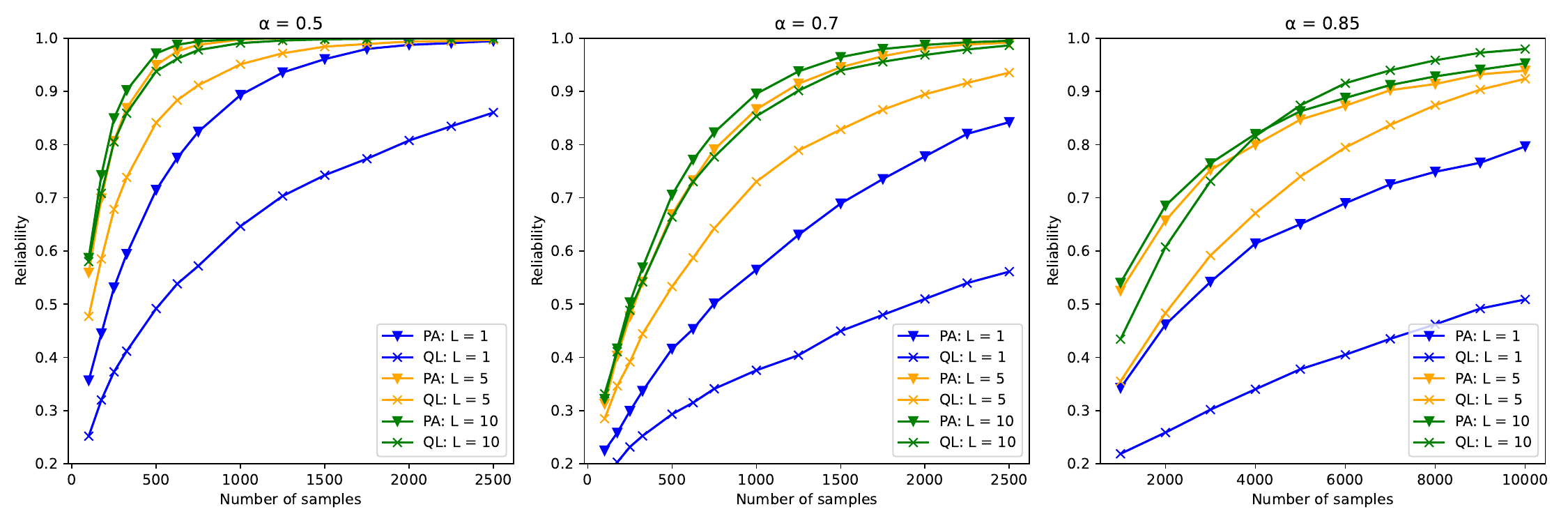}
    \caption{Reliability vs. number of samples for different $\alpha$ and $L$, with $K=10$ and $\rho=0.95$.  'PA' and 'QL' refer to Potential Actions and Queue Lengths, respectively. Less samples are needed for more substantial slowdowns to achieve the same level of reliability. The sampling interval $L$ has a profound impact on reliability if it is very small. The reliability when using marginal queue lengths is worse than when using potential actions in most cases, particularly for small values of $L$. }
    \label{fig:reliability}
\end{figure*}

The results are depicted in Figure \ref{fig:reliability}. We can see that less samples are needed for more substantial slowdowns to achieve the same level of reliability. We can also see that the sampling interval $L$ has a profound impact on reliability if it is very small. Intuitively, this is due to a high dependence between samples. Interestingly, this effect substantially diminishes as $L$ is slightly increased, suggesting that the dependence weakens very quickly with time. We remark that a practitioner can generate these plots and choose the values of $L$ and $N$ to achieve a desired level of reliability.

For comparison purposes, we also plot the reliability when using a marginal queue lengths based relative threshold test. As opposed to potential actions, it is not clear how to choose the threshold in this case. The test will look for a server whose queue length sample average is larger by a factor of $1+\gamma$ than the other queue length sample averages. In the event of a slowdown, given enough samples, these sample averages should be close to the steady state average queue lengths after slowdown $\EE\left[Q_1\right]$ (slow server) and $\EE\left[Q_2\right]$. Since we expect a slower server to have a larger average queue length, the intuitive choice would be to choose $\gamma$ so that $(1+\gamma)\EE\left[Q_2\right]$ is exactly half-way between $\EE\left[Q_1\right]$ and $\EE\left[Q_2\right]$. Therefore, we choose $\gamma$ so that
$(1+\gamma)\EE\left[Q_2\right]=(\EE\left[Q_1\right]+\EE\left[Q_2\right])/2.$

The results are depicted in Figure \ref{fig:reliability}. We can see that the reliability is worse than when using potential actions in most cases, particularly for small values of $L$, suggesting that the dependence of marginal queue lengths diminishes more slowly in time than that of potential actions.

\subsection{Reliability for Unknown Slowdown Factor}\label{subsec:sim_reliability_unknown}

The theoretical findings in this paper regarding the performance of the relative threshold test based on potential actions, as stated in Theorems \ref{thm:pa_detection_simple} and \ref{thm:normal_appox_reliability}, assume that the slowdown factor \( \alpha \) is known in advance, enabling the selection of an appropriate test threshold. However, in practice, slowdowns may occur with unforeseen magnitudes. This section aims to demonstrate that our approach remains effective even under such scenarios.

Specifically, one can choose a minimal slowdown magnitude of interest (e.g., 30\%), which corresponds to a maximal slowdown factor \( \alpha_{\text{max}} \) (e.g., \( 0.7 \)), and set the relative threshold factor to \( \gamma = (1 - 1.1\alpha_{\text{max}})/2 \). While a formal proof is not available, we provide numerical evidence that the test performs well even for actual slowdown factors smaller than \( \alpha_{\text{max}} \). The underlying intuition is that if the test is calibrated to detect a slowdown at \( \alpha_{\text{max}} \), smaller slowdowns should be easier to identify. However, this comes with a slight increase in false alarm rates when no slowdowns are present.

\begin{figure*}[h!]
    \centering
    \includegraphics[width=1\textwidth]{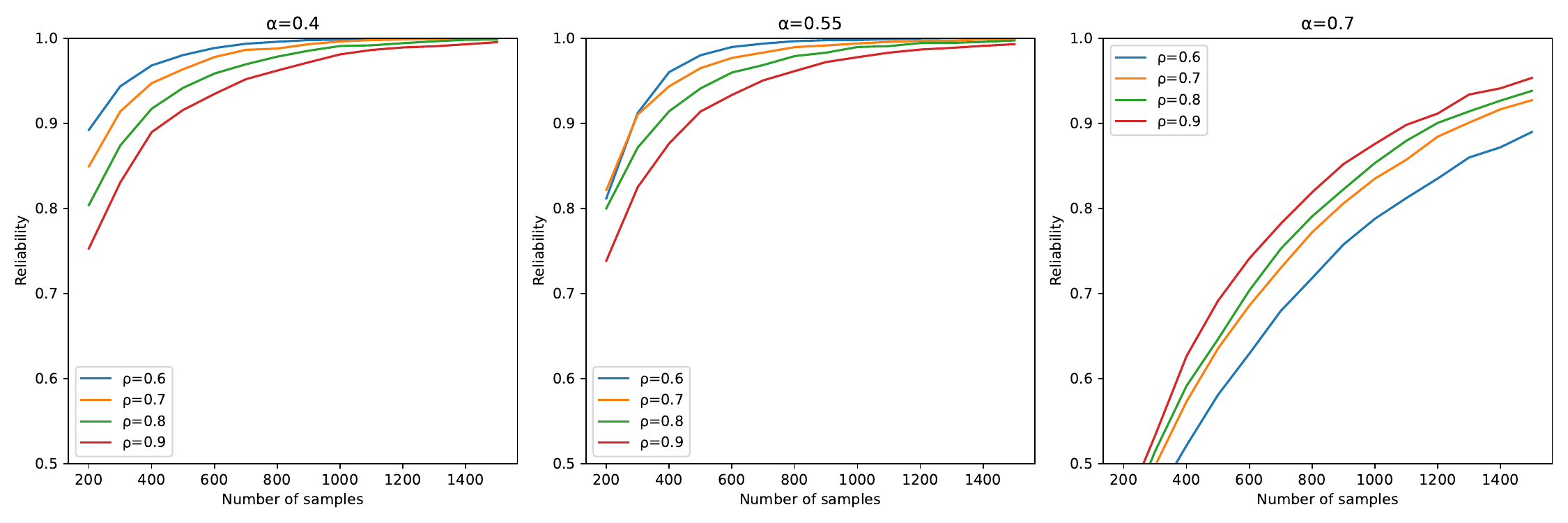}
    \caption{Reliability vs. number of samples for different values of $\alpha$ and $\rho$, where $K=10$, $L=10$ and $\alpha_{max}=0.7$. Even though the test parameters correspond to $\alpha_{max}$, the reliability for slowdowns with larger magnitudes is higher. }
    \label{fig:unknown}
\end{figure*}

To illustrate this, we conduct the following experiment. We consider \( K = 10 \) servers with a sampling interval of \( L = 10 \). Setting \( \alpha_{\text{max}} = 0.7 \), we adjust the test threshold accordingly and evaluate reliability across various slowdown magnitudes and system loads. Specifically, we test for \( \alpha \in \{0.4, 0.55, 0.7\} \) and \( \rho \in \{0.6, 0.7, 0.8, 0.9\} \), where \( \rho \) represents the load after the slowdown. The results are presented in Figure \ref{fig:unknown}.

We observe that the reliability is substantially higher for slowdown factors that are smaller than the one used to set the test parameters. Intuitively, if the actual slowdown is more substantial than what the test is designed to detect, then the probability of correct detection increases. This highlights the robustness and potential benefits of our approach when the slowdown magnitude is not known in advance.

Additionally, the results illustrate the significant impact of load (\( \rho \)) on reliability. In most cases, lower loads correspond to higher reliability. Intuitively, when \( L \) is fixed, higher loads increase the dependence between samples and reduce the signal-to-noise ratio, necessitating more samples to achieve comparable reliability.


\subsection{Example Use}\label{subsec:sim_example_use}

The purpose of this section is to illustrate how our results can be applied in practice. Specifically, we propose a straightforward procedure capable of inferring slowdowns in an online manner with high probability. The results can be displayed on a system operator's dashboard for real-time monitoring. Building on the numerical findings in Section \ref{subsec:sim_reliability_unknown}, we detail the procedure for cases where the slowdown magnitude is unknown in advance.

\noindent \textbf{Setup:}

\noindent \textbf{1. Specify Minimal Slowdown Magnitude: } 
   The operator first defines the minimal slowdown magnitude of interest, determining \( \alpha_{\text{max}} \) (e.g., detecting slowdowns of \( 30\% \) corresponds to \( \alpha_{\text{max}} = 0.7 \)). This choice reflects the operator's preferences, as a larger \( \alpha_{\text{max}} \) increases the likelihood of false alarms or necessitates more samples to achieve high reliability, potentially delaying detection.

\noindent \textbf{2. Set Test Parameters:}  
   The relative threshold test parameter is set to \( \gamma = (1 - 1.1\alpha_{\text{max}})/2 \). Based on simulation results (e.g., Figure \ref{fig:reliability}), the operator selects the number of samples \( N \) and sampling interval \( L \) to achieve a desired level of reliability. If \( L \) is predetermined, the operator can adjust \( N \) accordingly.

\noindent \textbf{Operation:}

As queue length samples arrive, the relative threshold test is applied using a sliding window of size \( N \). The test outputs 1 if it detects a slowdown and 0 otherwise. Before a slowdown, the test predominantly outputs 0s; after, it outputs mostly 1s. While informative, this can appear noisy.  

\begin{figure*}[h!]
    \centering
    \includegraphics[width=1\textwidth]{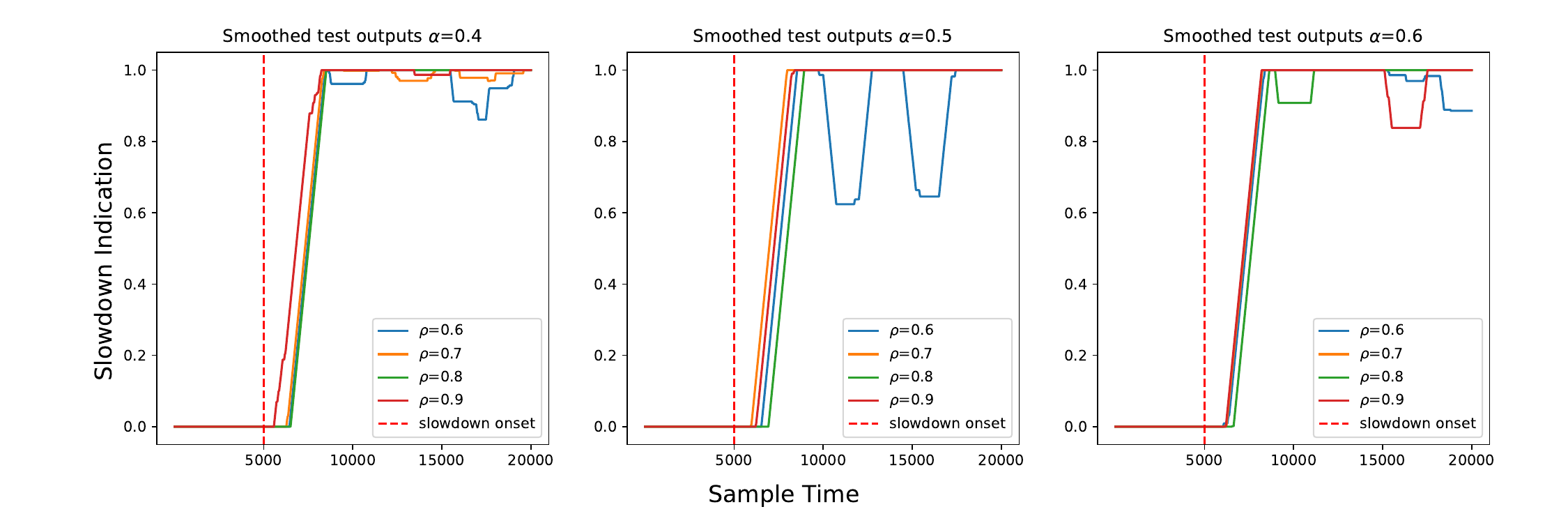}
    \caption{An Operator's dashboard, indicating whether there is a slowdown (y-axis values around 1) for different loads and slowdown factors.}
    \label{fig:dashboard}
\end{figure*}

To address this, a smoothing operation is applied, using a moving average of size \( N \) on the test output data. Figure \ref{fig:dashboard} illustrates the resulting dashboard for parameters \( K=10 \), \( L=10 \), \( \alpha_{\text{max}}=0.7 \), \( \alpha \in \{0.4, 0.5, 0.6\} \), \( \rho \in \{0.6, 0.7, 0.8, 0.9\} \), and \( N=2000 \). These parameters were chosen based on Figure \ref{fig:reliability}, where \( N=2000 \) is sufficient to achieve high reliability. The y-axis represents the smoothed test output in \([0, 1]\), while the x-axis corresponds to the sample index.

\noindent \textbf{Interpretation:}

In this example, a change occurs at time $50,000$. As $L=10$, this corresponds to sample $5,000$ in the x-axis. Before this point, all servers operate at the same processing speed, resulting in dashboard values of zero. After the change, server 1 experiences a slowdown, leading to a linear increase in the dashboard output, which stabilizes around 1. Occasional deviations below 1 after the change are due to stochasticity. Nevertheless, the dashboard quickly highlights potential issues, allowing operators to identify anomalies and initiate deeper investigations promptly.


\section{Concluding Remarks}

This paper takes a first step towards understanding how to best use observational data to perform effective slowdown learning in an effort to safeguard a service system's operational integrity. Using a novel stochastic and statistical framework, we uncover fundamental difficulties and subtleties of slowdown learning under adaptive congestion control, and propose a novel potential action statistic as an effective remedy.   We should highlight that the paper does not advocate for using the potential action as a sole statistic in designing a detection procedure (e.g., it would obviously be uninformative in non-adaptive systems where actions do not reflect underlying service disruptions). Instead, many machine-learning-based detection systems in practice often combine multiple sources of signals, and we believe that the potential action statistic proposed here can be used as one of the important information sources. Overall, we hope this work serves as a starting point for exploring the rich design space of statistical estimators for observational studies in dynamic service systems, as well as their intricate interactions with the underlying stochastic control mechanisms.

\vspace{10pt}


\bibliographystyle{apalike}
\bibliography{mybib} 

\begin{thebibliography}{}

\bibitem[Aronow and Samii, 2017]{arownow2017}
Aronow, P.~M. and Samii, C. (2017).
\newblock Estimating average causal effects under general interference, with application to a social network experiment.
\newblock {\em The Annals of Applied Statistics}, 11(4):1912--1947.

\bibitem[Asanjarani et~al., 2021]{asanjarani2021survey}
Asanjarani, A., Nazarathy, Y., and Taylor, P. (2021).
\newblock A survey of parameter and state estimation in queues.
\newblock {\em Queueing Systems}, 97(1):39--80.

\bibitem[Atar et~al., 2019]{atar2019replicate}
Atar, R., Keslassy, I., and Mendelson, G. (2019).
\newblock Replicate to the shortest queues.
\newblock {\em Queueing Systems}, 92:1--23.

\bibitem[Atar et~al., 2020]{atar2020persistent}
Atar, R., Keslassy, I., Mendelson, G., Orda, A., and Vargaftik, S. (2020).
\newblock Persistent-idle load-distribution.
\newblock {\em Stochastic Systems}, 10(2):152--169.

\bibitem[Bajari et~al., 2021]{bajari2021multiple}
Bajari, P., Burdick, B., Imbens, G.~W., Masoero, L., McQueen, J., Richardson, T., and Rosen, I.~M. (2021).
\newblock Multiple randomization designs.
\newblock {\em arXiv preprint arXiv:2112.13495}.

\bibitem[Baker et~al., 1991]{baker1991patients}
Baker, D.~W., Stevens, C.~D., and Brook, R.~H. (1991).
\newblock Patients who leave a public hospital emergency department without being seen by a physician: causes and consequences.
\newblock {\em Jama}, 266(8):1085--1090.

\bibitem[Bhat and Rao, 1987]{bhat1987statistical}
Bhat, U.~N. and Rao, S.~S. (1987).
\newblock Statistical analysis of queueing systems.
\newblock {\em Queueing Systems}, 1(3):217--247.

\bibitem[Borgs et~al., 2014]{borgs2014optimal}
Borgs, C., Chayes, J.~T., Doroudi, S., Harchol-Balter, M., and Xu, K. (2014).
\newblock The optimal admission threshold in observable queues with state dependent pricing.
\newblock {\em Probability in the Engineering and Informational Sciences}, 28(1):101--119.

\bibitem[Chen et~al., 1994]{chen1994parameter}
Chen, T.~M., Walrand, J., and Messerschmitt, D.~G. (1994).
\newblock Parameter estimation for partially observed queues.
\newblock {\em IEEE Transactions on communications}, 42(9):2730--2739.

\bibitem[Choudhury et~al., 2021]{choudhury2021job}
Choudhury, T., Joshi, G., Wang, W., and Shakkottai, S. (2021).
\newblock Job dispatching policies for queueing systems with unknown service rates.
\newblock In {\em Proceedings of the Twenty-second International Symposium on Theory, Algorithmic Foundations, and Protocol Design for Mobile Networks and Mobile Computing}, pages 181--190.

\bibitem[Clarke, 1957]{clarke1957maximum}
Clarke, A.~B. (1957).
\newblock Maximum likelihood estimates in a simple queue.
\newblock {\em The Annals of Mathematical Statistics}, 28(4):1036--1040.

\bibitem[Derlet, 2002]{derlet2002overcrowding}
Derlet, R.~W. (2002).
\newblock Overcrowding in emergency departments: increased demand and decreased capacity.
\newblock {\em Annals of emergency medicine}, 39(4):430--432.

\bibitem[Durrett, 2019]{durrett2019probability}
Durrett, R. (2019).
\newblock {\em Probability: theory and examples}, volume~49.
\newblock Cambridge university press.

\bibitem[Foss and Chernova, 1998]{foss1998stability}
Foss, S. and Chernova, N. (1998).
\newblock On the stability of a partially accessible multi-station queue with state-dependent routing.
\newblock {\em Queueing Systems}, 29:55--73.

\bibitem[Freund et~al., 2022]{freund2022efficient}
Freund, D., Lykouris, T., and Weng, W. (2022).
\newblock Efficient decentralized multi-agent learning in asymmetric queuing systems.
\newblock {\em arXiv preprint arXiv:2206.03324}.

\bibitem[Freund et~al., 2023]{freund2023quantifying}
Freund, D., Lykouris, T., and Weng, W. (2023).
\newblock Quantifying the cost of learning in queueing systems.
\newblock {\em arXiv preprint arXiv:2308.07817}.

\bibitem[Gallegos-Herrada et~al., 2024]{gallegos2024equivalences}
Gallegos-Herrada, M.~A., Ledvinka, D., and Rosenthal, J.~S. (2024).
\newblock Equivalences of geometric ergodicity of markov chains.
\newblock {\em Journal of Theoretical Probability}, 37(2):1230--1256.

\bibitem[Geyer, 1992]{geyer1992practical}
Geyer, C.~J. (1992).
\newblock Practical markov chain monte carlo.
\newblock {\em Statistical science}, pages 473--483.

\bibitem[Glynn and Whitt, 1992]{glynn1992asymptotic}
Glynn, P.~W. and Whitt, W. (1992).
\newblock The asymptotic validity of sequential stopping rules for stochastic simulations.
\newblock {\em The Annals of Applied Probability}, 2(1):180--198.

\bibitem[Haproxy, 2023]{HAproxyStarter2023}
Haproxy (2023).
\newblock Haproxy starter guide version 2.9-dev5-34.

\bibitem[Heckman, 2008]{heckman2008econometric}
Heckman, J.~J. (2008).
\newblock Econometric causality.
\newblock {\em International statistical review}, 76(1):1--27.

\bibitem[Hirano et~al., 2003]{hirano2003efficient}
Hirano, K., Imbens, G.~W., and Ridder, G. (2003).
\newblock Efficient estimation of average treatment effects using the estimated propensity score.
\newblock {\em Econometrica}, 71(4):1161--1189.

\bibitem[Hoeffding, 1994]{hoeffding1994probability}
Hoeffding, W. (1994).
\newblock Probability inequalities for sums of bounded random variables.
\newblock {\em The collected works of Wassily Hoeffding}, pages 409--426.

\bibitem[Hoot and Aronsky, 2008]{hoot2008systematic}
Hoot, N.~R. and Aronsky, D. (2008).
\newblock Systematic review of emergency department crowding: causes, effects, and solutions.
\newblock {\em Annals of emergency medicine}, 52(2):126--136.

\bibitem[Horvitz and Thompson, 1952]{horvitz1952generalization}
Horvitz, D.~G. and Thompson, D.~J. (1952).
\newblock A generalization of sampling without replacement from a finite universe.
\newblock {\em Journal of the American statistical Association}, 47(260):663--685.

\bibitem[Hudgens and Halloran, 2008]{hudgens_2008}
Hudgens, M.~G. and Halloran, M.~E. (2008).
\newblock Toward causal inference with interference.
\newblock {\em Journal of the American Statistical Association}, 103(482):832--842.

\bibitem[Ibidunmoye et~al., 2015]{ibidunmoye2015performance}
Ibidunmoye, O., Hern{\'a}ndez-Rodriguez, F., and Elmroth, E. (2015).
\newblock Performance anomaly detection and bottleneck identification.
\newblock {\em ACM Computing Surveys (CSUR)}, 48(1):1--35.

\bibitem[Ibragimov, 1962]{ibragimov1962some}
Ibragimov, I.~A. (1962).
\newblock Some limit theorems for stationary processes.
\newblock {\em Theory of Probability \& Its Applications}, 7(4):349--382.

\bibitem[Johari et~al., 2022]{johari2022experimental}
Johari, R., Li, H., Liskovich, I., and Weintraub, G.~Y. (2022).
\newblock Experimental design in two-sided platforms: An analysis of bias.
\newblock {\em Management Science}.

\bibitem[Jones, 2004]{jones2004markov}
Jones, G.~L. (2004).
\newblock On the markov chain central limit theorem.
\newblock {\em Probability Surveys}, 1:299--320.

\bibitem[Krishnasamy et~al., 2021]{krishnasamy2021learning}
Krishnasamy, S., Sen, R., Johari, R., and Shakkottai, S. (2021).
\newblock Learning unknown service rates in queues: A multiarmed bandit approach.
\newblock {\em Operations Research}, 69(1):315--330.

\bibitem[Larson, 1990]{larson1990queue}
Larson, R.~C. (1990).
\newblock The queue inference engine: Deducing queue statistics from transactional data.
\newblock {\em Management Science}, 36(5):586--601.

\bibitem[Li et~al., 2023]{li2023experimenting}
Li, S., Johari, R., Kuang, X., and Wager, S. (2023).
\newblock Experimenting under stochastic congestion.
\newblock {\em arXiv preprint arXiv:2302.12093}.

\bibitem[Lu et~al., 2011]{lu2011join}
Lu, Y., Xie, Q., Kliot, G., Geller, A., Larus, J.~R., and Greenberg, A. (2011).
\newblock Join-idle-queue: A novel load balancing algorithm for dynamically scalable web services.
\newblock {\em Performance Evaluation}, 68(11):1056--1071.

\bibitem[Luczak and McDiarmid, 2006]{luczak2006maximum}
Luczak, M.~J. and McDiarmid, C. (2006).
\newblock On the maximum queue length in the supermarket model.
\newblock {\em Annals of Probability}, 34(2):493--527.

\bibitem[Manski, 2013]{manski2013identification}
Manski, C.~F. (2013).
\newblock Identification of treatment response with social interactions.
\newblock {\em The Econometrics Journal}, 16(1):S1--S23.

\bibitem[Mendelson and Xu, 2022]{mendelson2022care}
Mendelson, G. and Xu, K. (2022).
\newblock Care: Resource allocation using sparse communication.
\newblock {\em arXiv preprint arXiv:2206.02410}.

\bibitem[Meyn and Tweedie, 1993]{meyn1993stability}
Meyn, S.~P. and Tweedie, R.~L. (1993).
\newblock Stability of markovian processes iii: Foster--lyapunov criteria for continuous-time processes.
\newblock {\em Advances in Applied Probability}, 25(3):518--548.

\bibitem[Meyn and Tweedie, 1994]{meyn1994computable}
Meyn, S.~P. and Tweedie, R.~L. (1994).
\newblock Computable bounds for geometric convergence rates of markov chains.
\newblock {\em The Annals of Applied Probability}, pages 981--1011.

\bibitem[Mitzenmacher, 2001]{mitzenmacher2001power}
Mitzenmacher, M. (2001).
\newblock The power of two choices in randomized load balancing.
\newblock {\em IEEE Transactions on Parallel and Distributed Systems}, 12(10):1094--1104.

\bibitem[Munro et~al., 2021]{munro2021treatment}
Munro, E., Wager, S., and Xu, K. (2021).
\newblock Treatment effects in market equilibrium.
\newblock {\em arXiv preprint arXiv:2109.11647}.

\bibitem[Patel et~al., 2006]{patel2006ambulance}
Patel, P.~B., Derlet, R.~W., Vinson, D.~R., Williams, M., and Wills, J. (2006).
\newblock Ambulance diversion reduction: the sacramento solution.
\newblock {\em The American journal of emergency medicine}, 24(2):206--213.

\bibitem[Pham et~al., 2006]{pham2006effects}
Pham, J.~C., Patel, R., Millin, M.~G., Kirsch, T.~D., and Chanmugam, A. (2006).
\newblock The effects of ambulance diversion: a comprehensive review.
\newblock {\em Academic Emergency Medicine}, 13(11):1220--1227.

\bibitem[Robert, 1995]{robert1995convergence}
Robert, C.~P. (1995).
\newblock Convergence control methods for markov chain monte carlo algorithms.
\newblock {\em Statistical science}, 10(3):231--253.

\bibitem[Robins et~al., 1994]{robins1994estimation}
Robins, J.~M., Rotnitzky, A., and Zhao, L.~P. (1994).
\newblock Estimation of regression coefficients when some regressors are not always observed.
\newblock {\em Journal of the American statistical Association}, 89(427):846--866.

\bibitem[Rosenbaum and Rubin, 1983]{rosenbaum1983central}
Rosenbaum, P.~R. and Rubin, D.~B. (1983).
\newblock The central role of the propensity score in observational studies for causal effects.
\newblock {\em Biometrika}, 70(1):41--55.

\bibitem[Rosenthal, 2002]{rosenthal2002quantitative}
Rosenthal, J. (2002).
\newblock Quantitative convergence rates of markov chains: A simple account.

\bibitem[Savva and Tezcan, 2019]{savva2019reduce}
Savva, N. and Tezcan, T. (2019).
\newblock To reduce emergency room wait times, tie them to payments.
\newblock {\em Harvard Business Review.}

\bibitem[Savva et~al., 2019]{savva2019can}
Savva, N., Tezcan, T., and Y{\i}ld{\i}z, {\"O}. (2019).
\newblock Can yardstick competition reduce waiting times?
\newblock {\em Management Science}, 65(7):3196--3215.

\bibitem[Shah and Prabhakar, 2002]{shah2002use}
Shah, D. and Prabhakar, B. (2002).
\newblock The use of memory in randomized load balancing.
\newblock In {\em Proceedings IEEE International Symposium on Information Theory,}, page 125. IEEE.

\bibitem[Shi et~al., 2019]{shi2019process}
Shi, C., Wei, Y., and Zhong, Y. (2019).
\newblock Process flexibility for multiperiod production systems.
\newblock {\em Operations Research}, 67(5):1300--1320.

\bibitem[Shirazi et~al., 2017]{shirazi2017extended}
Shirazi, S.~N., Gouglidis, A., Farshad, A., and Hutchison, D. (2017).
\newblock The extended cloud: Review and analysis of mobile edge computing and fog from a security and resilience perspective.
\newblock {\em IEEE Journal on Selected Areas in Communications}, 35(11):2586--2595.

\bibitem[Stidham, 1985]{stidham1985optimal}
Stidham, S. (1985).
\newblock Optimal control of admission to a queueing system.
\newblock {\em IEEE Transactions on Automatic Control}, 30(8):705--713.

\bibitem[Stolyar, 2004]{stolyar2004maxweight}
Stolyar, A.~L. (2004).
\newblock Maxweight scheduling in a generalized switch: State space collapse and workload minimization in heavy traffic.
\newblock {\em The Annals of Applied Probability}, 14(1):1--53.

\bibitem[Varma and Maguluri, 2021]{varma2021transportation}
Varma, S.~M. and Maguluri, S.~T. (2021).
\newblock Transportation polytope and its applications in parallel server systems.
\newblock {\em arXiv preprint arXiv:2108.13167}.

\bibitem[Vvedenskaya et~al., 1996]{vvedenskaya1996queueing}
Vvedenskaya, N.~D., Dobrushin, R.~L., and Karpelevich, F.~I. (1996).
\newblock Queueing system with selection of the shortest of two queues: An asymptotic approach.
\newblock {\em Problemy Peredachi Informatsii}, 32(1):20--34.

\bibitem[Wager and Xu, 2021]{wager2021experimenting}
Wager, S. and Xu, K. (2021).
\newblock Experimenting in equilibrium.
\newblock {\em Management Science}, 67(11):6694--6715.

\bibitem[Walton and Xu, 2021]{walton2021learning}
Walton, N. and Xu, K. (2021).
\newblock Learning and information in stochastic networks and queues.
\newblock In {\em Tutorials in Operations Research: Emerging Optimization Methods and Modeling Techniques with Applications}, pages 161--198. INFORMS.

\bibitem[Wang et~al., 2011]{wang2011statistical}
Wang, C., Viswanathan, K., Choudur, L., Talwar, V., Satterfield, W., and Schwan, K. (2011).
\newblock Statistical techniques for online anomaly detection in data centers.
\newblock In {\em 12th IFIP/IEEE international symposium on integrated network management (IM 2011) and workshops}, pages 385--392. IEEE.

\bibitem[Whitt, 2002]{whitt2002stochastic}
Whitt, W. (2002).
\newblock Stochastic-process limits: an introduction to stochastic-process limits and their application to queues.
\newblock {\em Space}, 500:391--426.

\bibitem[Wolff, 1982]{wolff1982poisson}
Wolff, R.~W. (1982).
\newblock Poisson arrivals see time averages.
\newblock {\em Operations research}, 30(2):223--231.

\bibitem[Xu and Chan, 2016]{xu2016using}
Xu, K. and Chan, C.~W. (2016).
\newblock Using future information to reduce waiting times in the emergency department via diversion.
\newblock {\em Manufacturing \& Service Operations Management}, 18(3):314--331.

\bibitem[Yadav, 2023]{Yadav_2023}
Yadav, G. (2023).
\newblock Haproxy monitoring guide: Important metrics; best tools.

\bibitem[Ye, 2021]{ye2021optimal}
Ye, H.-Q. (2021).
\newblock Optimal routing to parallel servers in heavy traffic.
\newblock {\em Available at SSRN 3996615}.

\bibitem[Zhong et~al., 2022]{zhong2022learning}
Zhong, Y., Birge, J.~R., and Ward, A. (2022).
\newblock Learning the scheduling policy in time-varying multiclass many server queues with abandonment.
\newblock {\em Available at SSRN}.

\end{thebibliography}


\newpage

\setcounter{page}{1} 

\begin{center}
    {\Large Electronic Companion - Learning Service Slowdown Using Observational Data}
\end{center}

\begin{APPENDICES}

\section{Proofs}

\subsection{Proof of Theorem \ref{thm:no_ql_detection}}
\label{app:thm:no_ql_detection}

\noindent \textbf{Theorem \ref{thm:no_ql_detection}}. Consider the problem setting in Definition \ref{def:known_slowdown}. Fix any admissible marginal statistic $g\in\mathcal{G}$ and a relative threshold decision rule satisfying Definition \eqref{def:rel_threshold}. Suppose the congestion control policy is JSQ. Then for {any} $\alpha \in (0,1)$ and $N \in \mathbb{N}$, the reliability vanishes in heavy traffic: 
\begin{equation*}
	p_+  = \min_{C \in \{C1, C2\} } \pb_C(\mbox{success}) \to 0, \quad \mbox{as } \lambda \to \alpha + (K-1). 
\end{equation*}

\emph{Preliminaries.}  Note that whenever the system is under-loaded with $\rho<1$, the JSQ congestion control policy stabilizes the system in that the resulting queue length process $Q(\cdot)$ is positive recurrent \citep{foss1998stability}. 

We prove a slightly stronger version of the theorem, where we only require the congestion control policy to be maximally stable (Definition \ref{def:maxStable}) and to satisfy the State-Space-Collapse property \citep{stolyar2004maxweight} defined below.  
\begin{definition}[State Space Collapse (SSC)]
\label{def:SSC}
 Let $Q = (Q_1, \ldots, Q_K)$ be distributed according to the steady-state distribution of $\{Q(t)\}_{t\in \rp}$. We say that the congestion control policy induces state space collapse if
$$(1-\rho)(Q_1,\ldots,Q_K) \xRightarrow{d} (X,\ldots,X),$$
where $X$ is a non-trivial continuous random variable with finite first and second moments, and the limit of $\rho \to 1$ is taken by fixing the service rates while letting $\lambda$ approach the sum of the service rates from below. 
\end{definition}
State space collapse is a property that is often associated with superior delay performance \citep{shi2019process, varma2021transportation}. Policies satisfying SSC generally aim to direct incoming jobs towards less congested servers (smaller queues). In doing so, they tend to minimize system-wide delay by preventing long queues from building up.  The Join-the-Shortest-Queue (JSQ) policy satisfies SSC \citep{ye2021optimal}, and so does a communication-efficient variants of JSQ that dispatch based on approximate queue length states \citep{mendelson2022care}. Another example is the Replicate-to-the-Shortest-Queues (RSQ($d$)), which employs job replication and subsequent adaptive cancellations in an effort to reduce delay  \citep{atar2019replicate}. 
\footnote{Strictly speaking, the SSC property for approximate JSQ and RSQ were established for a transient version of SSC. It is however generally believed that a steady-state SSC in the form we consider here holds for these policies as well.  The maximum stability for RSQ($d$) is shown for $d=1$ or $K$, and is conjectured to hold for $1 < d< K$ \citep{atar2019replicate}. }

\emph{Proof of Theorem \ref{thm:no_ql_detection}.} The high-level intuition for the proof is as follows. First, because the reliability is defined to be the minimum of success probabilities across different underlying states, in order to prove a negative result on reliability, it suffices to consider C1 only. Under C1, and using the state space collapse property, we argue that the marginal queue length distribution across different servers converge (up to a load-dependent factor) to the same limiting distribution. By the weak law of large numbers and the continuous mapping theorem, it then further implies that the marginal statistics will necessarily be very similar across the different servers, leading to the failure of any relative threshold rule. 

To make the limit $\lambda \to \alpha + K+1$ more explicit for analysis, we will consider the following sequence: 
\begin{equation*}
    \rho_m = 1-1/\sqrt{m}, \mbox{ and } \lambda_m =  (\alpha + K-1)\rho_m, \quad m \in \N. 
\end{equation*}
For a given $m$, we will denote the by $\hat{Q}_{k,n}$ the scaled queue length sample
\begin{equation*}
    \hat Q^m_{k,n}:=(1-\rho_m)Q_{k,n} = \frac{1}{\sqrt{m}}Q_{k,n}, 
\end{equation*}
where for simplicity of notation we will suppress the dependence on $m$ in $\hat Q^m_{k,n}$ in the remainder of the proof. In particular, the vector $\hat Q_{k,\cdot}=\p{\hat Q_{k,1}, \cdots, \hat Q_{k, N}}$ represents the scaled marginal queue length data collected from server $k$. 

We will prove the claim by analyzing the limiting distribution of the sequence of the scaled marginal queue length vectors $\hat{Q}_{k, \cdot}$ as $ m \to \infty$. The SSC property (Definition \ref{def:SSC}) ensures that 
\begin{equation*}
    \hat Q_{\cdot,n} \stackrel{d}{\Rightarrow} (X, X, \ldots, X), \quad \mbox{as $m \to \infty$.}
\end{equation*}
Furthermore, because the $Q_{\cdot,n}$'s are sampled i.i.d.~across $n$, we have that for all $k$, the scaled marginal queue length vector vector $\hat Q_{k,\cdot}$ satisfies
\begin{equation}
    \hat Q_{k,\cdot} \stackrel{d}{\Rightarrow} (X_1, X_2, \ldots, X_N), \quad \mbox{as $m \to \infty$,}
    \label{eq:SSC sync per server} 
\end{equation}
where the $X_i$'s are i.i.d.~and distributed according to $X$.  To summarize, letting $\bar X = (X_1, X_2, \ldots, X_N)$, we have that
\begin{equation}\label{eq:SSC sync}  
(\hat Q_{1, \cdot}, \hat Q_{2, \cdot}, \ldots, \hat Q_{K, \cdot}) \stackrel{d}{\Rightarrow}  \p{\bar X, \bar X, \ldots, \bar X}. 
\end{equation}
In particular, for any pair of servers $k$ and $j$, 
\begin{equation}
\label{eq:SSC sync pair of servers} 
(\hat{Q}_{k, \cdot},\hat{Q}_{j, \cdot}) \stackrel{d}{\Rightarrow} (\bar X,\bar X).
\end{equation}

We now consider the success probability under a majorizing threshold rule; the same proof applies to a minorizing threshold rule as well due to the equivalence between the two (Remark \ref{rem:majorminor_thresh}). Fix $\gamma >0$. Recall that we have assumed that server $1$ has slowed down while all other servers remain healthy, and a successful detection corresponds to the decision rule outputting 1. We have: 
\begin{align}\label{eq:jsq ss1}
    \pb_{C1}(\mbox{success}) =& \pb\Big(g(Q_{1,\cdot})\geq (1+\gamma)\max_{j\neq 1}g(Q_{j, \cdot})\Big) \nln
    = & \pb\Big(\frac{1}{\sqrt{m}}g(Q_{1,\cdot})\geq (1+\gamma)\max_{j\neq 1}\frac{1}{\sqrt{m}}g(Q_{j, \cdot})\Big) \nln
    \sk{(a)}{=}& \pb\Big(g(\hat{Q}_{1,\cdot})\geq (1+\gamma)\max_{j\neq 1}g(\hat{Q}_{j,\cdot})\Big) \nln
    \leq& \pb\Big(g(\hat{Q}_{1,\cdot})\geq (1+\gamma)g(\hat{Q}_{2,\cdot})\Big) \nln
    = & \pb\Big(g(\hat{Q}_{1,\cdot})-g(\hat{Q}_{2,\cdot})\geq \gamma g(\hat{Q}_{2,\cdot})\Big),
\end{align}
where (a) follows from the scale invariance of $g$. 

The remainder of the proof will focus on showing that the probability of the event 
\begin{equation*}
    \{g(\hat{Q}_{1,\cdot})-g(\hat{Q}_{2,\cdot})\geq \gamma g(\hat{Q}_{2,\cdot})\}
\end{equation*}
will vanish as $m$ increases. By the SSC property and the continuity of $g$, we expect the difference on the left-hand side of the inequality to converge to zero in probability, while the right-hand side to converge to a non-trivial, non-negative random variable. Therefore, one would expect the probability of this event to vanish as $m$ increases, which will conclude the proof. 

To make the above argument concrete, fix $\epsilon_1>0$ whose value depends on $\delta$ and will be chosen later. We have:
\begin{align}\label{eq:jsq ss2}
    & \pb\Big(g(\hat{Q}_{1,\cdot})-g(\hat{Q}_{2,\cdot})\geq \gamma g(\hat{Q}_{2,\cdot})\Big) \nln
    = & 
    \pb\Big(g(\hat{Q}_{1,\cdot})-g(\hat{Q}_{2,\cdot})\geq \gamma g(\hat{Q}_{2,\cdot}),g(\hat{Q}_{2,\cdot})>\epsilon_1\Big)
    + 
    \pb\Big(g(\hat{Q}_{1,\cdot})-g(\hat{Q}_{2,\cdot})\geq \gamma g(\hat{Q}_{2,\cdot}),g(\hat{Q}_{2,\cdot})\leq\epsilon_1\Big)\cr
    \leq & \pb\Big(g(\hat{Q}_{1,\cdot})-g(\hat{Q}_{2,\cdot})\geq \gamma\epsilon_1\Big)+\pb\Big(g(\hat{Q}_{2,\cdot})\leq\epsilon_1\Big).
\end{align}
Thus, combining \eqref{eq:jsq ss1} and \eqref{eq:jsq ss2} yields:
\begin{align}\label{eq:jsq main}
    \pb_{C1}(\mbox{success})\leq \pb\Big(g(\hat{Q}_{1,\cdot})-g(\hat{Q}_{2,\cdot})\geq \gamma\epsilon_1\Big)+\pb\Big(g(\hat{Q}_{2,\cdot})\leq\epsilon_1\Big).
\end{align}

We now show that we can choose $\epsilon_1$ to be small enough such that for all large enough $m$, each of the two terms on the right-hand side of \eqref{eq:jsq main} is at most $\delta/2$.  Starting with $\pb\Big(g(\hat{Q}_{2,\cdot})\leq\epsilon_1\Big)$, by the SSC property (Definition \ref{def:SSC}) we have the convergence given in \eqref{eq:SSC sync per server}, namely that $\hat{Q}_{2,\cdot}\Rightarrow (X_1,\ldots,X_N)$. Since $g$ is continuous, by the continuous mapping theorem \citep[cf.][Theorem 3.2.11]{durrett2019probability}, we also have that 
\begin{equation*}
    g(\hat{Q}_{2,\cdot})\Rightarrow g(X_1,\ldots,X_N). 
\end{equation*}
By Portemanteau's theorem \citep[cf.][Theorem 3.2.11]{durrett2019probability}, using our assumption that $X$ is a continuous random variable, this implies that $\pb\Big(g(\hat{Q}_{2,\cdot})\leq\epsilon_1\Big)\xrightarrow{m \rightarrow \infty}\pb\Big(g(X_1,\ldots,X_N)\leq\epsilon_1\Big)$, meaning, that there exists $m_1(\epsilon_1)>0$ such that $\forall m>m_1(\epsilon_1)$ we have:
\begin{equation}\label{eq:jsq ss3}
    \pb\Big(g(\hat{Q}_{2,\cdot})\leq\epsilon_1\Big)\leq\pb\Big(g(X_1,\ldots,X_N)\leq\epsilon_1\Big)+\delta/4.
\end{equation}
Since $g(X_1,\ldots,X_N)$ is a continuous, non-negative  random variable, by the continuity of probability measures, we can choose $\epsilon_1$ to be sufficiently small such that 
\begin{equation}\label{eq:jsq ss4}
    \pb\Big(g(X_1,\ldots,X_N)\leq\epsilon_1\Big)\leq \delta/4.
\end{equation}
Combining \eqref{eq:jsq ss3} and \eqref{eq:jsq ss4} yields:
\begin{equation}\label{eq:jsq ss5}
    \pb\Big(g(\hat{Q}_{2,\cdot})\leq\epsilon_1\Big)\leq\delta/2.
\end{equation}


We now turn to $\pb\Big(g(\hat{Q}_{1,\cdot})-g(\hat{Q}_{2,\cdot})\geq \gamma\epsilon_1\Big)$. 

Define the function $h(x,y)=g(x)-g(y)$. Since $g$ is continuous, so is $h$. By the SSC property, we have the convergence given in \eqref{eq:SSC sync pair of servers}, namely, $(\hat{Q}_1,\hat{Q}_{2,\cdot})\Rightarrow (X_1,\ldots X_N,X_1,\ldots,X_N)$. By the continuous mapping theorem, we have that 
\begin{equation*}   h(\hat{Q}_1,\hat{Q}_{2,\cdot})\Rightarrow h(\bar X,\bar X) =g(\bar X)-g(\bar X )=0, 
\end{equation*}
which implies that $h(\hat{Q}_1,\hat{Q}_{2,\cdot})\rightarrow 0$ in probability. By the definition of convergence in probability, there exists $m_2(\gamma,\epsilon_1)>0$ such that $\forall m>m_2(\gamma,\epsilon_1)$ we have:
\begin{equation}\label{eq:jsq ss6}
    \pb\Big(g(\hat{Q}_{1,\cdot})-g(\hat{Q}_{2,\cdot})\geq \gamma\epsilon_1\Big)=\pb\Big(h(\hat{Q}_1,\hat{Q}_{2,\cdot})\geq \gamma\epsilon_1\Big)\leq \delta/2.
\end{equation}

To complete the proof, let us summarize how the various constants are chosen and in what order. Recall that $\gamma$ and $\delta$ are fixed. We then choose $\epsilon_1$ which depends on $\delta$, such that \eqref{eq:jsq ss4} holds. The value of $m_1(\epsilon_1)$ is then chosen such that \eqref{eq:jsq ss3} holds, which, together with \eqref{eq:jsq ss4} implies that \eqref{eq:jsq ss5} holds. The value of $m_2(\gamma,\epsilon_1)$ is then chosen so that 
\eqref{eq:jsq ss6} holds. Denoting $m_0=\max\{m_1,m_2\}$, we obtain that $\forall m> m_0$, we have:
\begin{align*}
    p_+ \leq \pb_{C1}(\mbox{success})\leq \pb\Big(g(\hat{Q}_{1,\cdot})-g(\hat{Q}_{2,\cdot})\geq \gamma\epsilon_1\Big)+\pb\Big(g(\hat{Q}_{2,\cdot})\leq\epsilon_1\Big)\leq \delta,
\end{align*}
where the first inequality is due to \eqref{eq:jsq main}, and the second inequality is due to \eqref{eq:jsq ss5} and \eqref{eq:jsq ss6}. That is, we have shown that for any $\delta>0$, the probability of success would be less than $\delta$ for all sufficiently large $m$, that is, all   $\lambda$ sufficiently close to $\alpha + K-1$. This completes the proof of Theorem \ref{thm:no_ql_detection}. \qed


\subsection{Proof of Lemma \ref{lemma:ub and lb on ratio}}
\label{app:lemma:ub and lb on ratio}

\noindent \textbf{Lemma \ref{lemma:ub and lb on ratio}.} 
  Fix $\alpha \in (0,1]$. Suppose C1 is true, and $\alpha <\lambda<\alpha+K-1$. Then under any maximally stable and symmetric control policy, the following holds. 
   \begin{enumerate}
       \item Let $\lambda_k$ be the stationary arrival rate of jobs to server $k$. Then
       \begin{equation}\label{eq:app:lemm1:1}
            \pi^R_k = \lambda_k/\lambda, \quad k = 1, \ldots, K. 
        \end{equation}
        \item $\pi^R_k=\pi^R_2$ for all $k>1$ and 
\begin{equation}
    \alpha\left(1 -(1-\rho)\frac{K-1+\alpha}{\alpha}\right) \leq \frac{\pi^R_1}{\pi^R_2} \leq \alpha \left ( \frac{1}{\rho-(1-\rho)\frac{\alpha}{K-1}}\right).
    \label{eq:app:lemm1:2}
\end{equation}
\item As $\rho \to 1$ (i.e., $\lambda \to \alpha+K-1$), we have
\begin{equation*}
      \pi^R_1 \to  \frac{\alpha}{K-1+\alpha},  \quad    \pi^R_2 \to \frac{1}{K-1+\alpha}, \quad \frac{\pi^R_1}{\pi^R_2} \to \alpha. \\
\end{equation*}
   \end{enumerate}

\noindent \textbf{Proof.}  For the first claim, recall that $\rho <1$ and the congestion control policy is maximally stable. Then $Q(\cdot)$ is positive recurrent and the $\lambda_k$'s are well defined. \eqref{eq:app:lemm1:1} follows from the Poisson Arrivals See Time Averages (PASTA) property \citep{wolff1982poisson}: because the arrival processes are Poisson and are independent of all other parts of the system, the long-run fraction of arrivals seeing $Q(\cdot)$ in a certain state $q$ is equal to the long-run fraction of time $Q(t) = q$. Because the potential action $\hat R_k(q)$ is defined to be the probability that an arrival would join queue $k$ upon seeing the queue lengths $q$, we have
\begin{equation*}
    \lambda_k = \lambda \sum_{q} \hat R_k(q) \pb(Q(0)=q) = \lambda \EE\left[\hat R_k(Q(0))\right]   = \lambda \pi^R_k. 
\end{equation*}

For the second claim, recall that under C1, we have $\mu_1 = \alpha$, $\mu_k=1$ for all $k>1$. Under a symmetric policy, it is clear that the steady-state potential action for a server only depends on whether it is healthy. Therefore, $\pi^R_2=\pi^R_k$ for all $k\geq 2$.   Because the system is stable, we must have that $\lambda_k < \mu_k$ for all $k$. By \eqref{eq:app:lemm1:1}, we thus have: 
	\begin{align}
		\label{eq:ps_upper}
	 \pi^R_1=  \lambda_1/\lambda \leq &\alpha / \lambda = \frac{\alpha}{\rho (K-1+\alpha)},  \\
	 \pi^R_2=  \lambda_2/\lambda = & 1/\lambda = \frac{1}{\rho (K-1+\alpha)}. 
	 		\label{eq:ph_upper}
	\end{align}
For the lower bound in \eqref{eq:app:lemm1:2}, we have
 \begin{align*}
     \frac{\pi^R_1}{\pi^R_2}\overset{(a)}{=}& \frac{1-(K-1)\pi^R_2}{\pi^R_2} \nln
     =& \frac{1}{\pi^R_2}-(K-1)  \nln
     \overset{(b)}{\geq}& \rho (K-1+\alpha)-(K-1) \nln
     = & \alpha\left(1 -(1-\rho)\frac{K-1+\alpha}{\alpha}\right) 
 \end{align*}
where $(a)$ follows from the fact that $\pi^R_k=\pi^R_2$ for all $k\geq 2$, $\sum_i \pi^R_i=1$, and therefore: 
	\begin{equation}
		\pi^R_1 + (K-1)\pi^R_2=1,
    \label{eq:prob sum}
	\end{equation}
and $(b)$ from \eqref{eq:ph_upper}. Similarly, for the upper bound in \eqref{eq:app:lemm1:2}, we have
 \begin{align*}
     \frac{\pi^R_1}{\pi^R_2} \overset{(a)}{=} & (K-1)\frac{\pi^R_1}{1-\pi^R_1}  \nln
     \overset{(b)}{\leq}& (K-1)\frac{ \frac{\alpha}{\rho (K-1+\alpha)}}{1- \frac{\alpha}{\rho (K-1+\alpha)}} \nln
     = & \alpha\left( \frac{K-1}{\rho (K-1+\alpha)-\alpha}\right) \nln
     = & \alpha \left ( \frac{1}{\rho-(1-\rho)\frac{\alpha}{K-1}}\right),
 \end{align*}
 where $(a)$ follows from \eqref{eq:prob sum} and $(b)$ from \eqref{eq:ps_upper}. 
 
 For the last claim, the upper and lower bounds above immediately imply that $\lim_{\rho \rightarrow 1}\frac{\pi^R_1}{\pi^R_2}=\alpha$. Combining \eqref{eq:prob sum} and \eqref{eq:ps_upper} yields:
	\begin{equation}
		\pi^R_2\geq  \frac{1-\frac{\alpha}{\rho (K-1+\alpha)} }{K-1}. 
		\label{eq:ph_lower}
	\end{equation}
Taking the limit as $\rho \to 1$ in \eqref{eq:ph_upper}  and \eqref{eq:ph_lower}  yields that
\begin{equation*}
	\lim_{\rho \to 1} \pi^R_2 = \frac{1}{K-1+\alpha}.
\end{equation*}
Repeating the same argument but for $\pi^R_1$ yields that
	\begin{equation*}
		\lim_{\rho \to 1} \pi^R_1 = \frac{\alpha}{K-1+\alpha}. 
	\end{equation*}
 This completes the proof.  \qed


\subsection{Proof of Theorem \ref{thm:pa_detection_simple}}
\label{app:thm:pa_detection_simple}

\noindent \textbf{Theorem \ref{thm:pa_detection_simple}}. 
Consider the problem setting in Definition \ref{def:known_slowdown}.  Suppose the congestion control policy is symmetric and maximally stable. Fix the summary statistics to be that of potential action and the decision rule a minorizing relative threshold rule with 
 \begin{equation*}
     \gamma=\frac{1-\alpha}{2}.
 \end{equation*} 
 Then, for all $N >0$ 
	\begin{equation*}
		      p_+ \geq 1- \max \left\{ (K-1) \exp\p{-\frac{(1-\alpha)^2}{32K^2}N}, \, K\exp\p{-\frac{(1-\alpha)^2}{8K^2}N}  \right\} \nonumber,  
	\end{equation*}
 as  $\lambda \to \alpha + K-1$. \\

\noindent \textbf{Proof.} We will use the following version of Hoeffding's inequality: 
\begin{lemma}[\citep{hoeffding1994probability}]
\label{lemma: hoeffding_app_general}
	Let $X_1,\ldots,X_N$ be independent random variables such that $a\leq X_i \leq b$ for all $i$, and let $S_N:=X_1+\ldots X_N$. Then, for all $t>0$, We have:
	\begin{equation*}
		\mathbb{P}(S_N-\mathbb{E}\left[S_N\right]\geq Nt)\leq e^{-\frac{2t^2N}{(b-a)^2}}.
	\end{equation*}
\end{lemma}
 
Define:
\begin{equation*}
	Y_k := \sum_{t=1}^N \hat R_{n,k} = N h_k(\calQ) , \quad k \in \{1, \ldots, K\}. 
\end{equation*}
Then, the relative threshold test would output the index $k$ if 
\begin{equation*}
{Y_k}< (1-\gamma) Y_j, \quad \forall j \neq k, 
\end{equation*}
and $\emptyset$, otherwise.  Define the events 
\begin{equation}\label{EC:eq:iid_pa_E_kj_events}
	E_{kj} = \left\{Y_k < (1-\gamma)Y_j \right\}, \quad  k,j \in \{1, \ldots, K\}. 
\end{equation}

Now, suppose the hypothesis C1 is true, so that  server 1 is slowed down. In this case, $\cap_{j \neq 1} E_{1j}$ corresponds to success event for the relative threshold test by correctly outputting 1. We now derive a lower bound on its probability. 
Using a union bound, we have that
\begin{align}\label{eq:proof_thm_2_C1}
\pb_{C1}(\mbox{success}) = & \pb(\cap_{j \neq 1} E_{1j}) = 1- \pb(\cup_{j \neq 1} \overline E_{1j}) \geq 1- \sum_{j\neq 1}\pb(\overline  E_{1j})\overset{(a)}{=}1- (K-1)\pb(\overline  E_{12} )\cr
=& 1- (K-1)\pb(Y_1 \geq (1-\gamma)Y_2),
\end{align}
where equality (a) is due to the symmetry between the healthy servers. 

Recall that the columns of $\calQ$ are sampled i.i.d.~from the stationary distribution of $Q(\cdot)$.  We have
\begin{align*}
    \pb(Y_1 \geq  (1 - \gamma)Y_2)=\pb(Y_1 - (1-\gamma)Y_2 \geq  0)=
    \pb \left(\sum_{i=1}^N \left( \hat R_{i,1}-(1-\gamma) \hat R_{i, 2}\right)\geq 0 \right)
\end{align*}

Let $W_i := \hat R_{i,1}-(1-\gamma) \hat R_{i, 2}$. Then $\{W_i\}$ is a sequence of bounded i.i.d. random variables, with   
\begin{equation*}
    W_i \in [-1, 1], \quad \EE\left[W_i\right]=\pi^R_1-(1-\gamma)\pi^R_2.
\end{equation*}
By \ref{lemma:ub and lb on ratio}, for all sufficiently large $\lambda$, we have that 
\begin{equation*}
    \pi^R_1 / \pi^R_2 < \alpha + \frac{1-\alpha}{4}, 
\end{equation*} 
and $\pi^R_2 >1/K$.
Recalling that $\gamma = (1-\alpha)/2$, we have that
\begin{align*}
    \EE\left[W_i\right]= & \pi^R_1-(1-\gamma) \pi^R_2 \nln
    =& \pi^R_2\left( \frac{\pi^R_1}{\pi^R_2}-(1-\gamma) \right) \nln
    <& \pi^R_2\left( \alpha+\frac{1-\alpha}{4}- \frac{\alpha+1}{2}\right)\nln
    = & - \pi^R_2 \frac{1-\alpha}{4}\nln
    < & - \frac{1-\alpha}{4K}. 
\end{align*}
Thus, using the Hoeffding inequality (Lemma \ref{lemma: hoeffding_app_general}), we have that 
\begin{align}\label{eq:thm_2_proof_last_C1}
\pb(Y_1 \geq (1-\gamma)Y_2) &=\pb\p{ \sum_{i=1}^N W_i-N\EE\left[W_i\right]\geq-N\EE\left[W_i\right]} \nln
\leq & \exp\p{-\frac{2(\EE\left[W_i\right])^2N}{4}} \nln
\leq & \exp\p{-\frac{(1-\alpha)^2}{32K^2}N}. 
\end{align}
Combining \eqref{eq:thm_2_proof_last_C1} with \eqref{eq:proof_thm_2_C1}, we have that 
\begin{align}
    \pb_{C1}(\mbox{success}) \geq  1-(K-1)\exp\p{-\frac{(1-\alpha)^2}{32K^2}N}.
    \label{eq:C1_succ_iid}
\end{align}

Now, suppose C2 is true and all servers are identical and run at unit rate and by symmetry, $\pi^R_1 = \pi^R_2 = 1/K$. In this case, success corresponds to the rule outputting $\emptyset$, which, using the events $E_{kj}$ defined in \eqref{EC:eq:iid_pa_E_kj_events}, corresponds to the event $\cap_{k}\cup_{j\neq k}\overline{E}_{kj}$. We have
\begin{align}\label{eq:proof_thm_2_C2}
\pb_{C2}(\mbox{success}) = & \pb(\cap_{k}\cup_{j\neq k}\overline{E}_{kj}) = 1- \pb(\cup_{k}\cap_{j\neq k}E_{kj}) \geq 1- \sum_{k}\pb(\cap_{j\neq k}E_{kj})=1- K\pb(\cap_{j\neq k}E_{1j} )\cr
\geq & 1- K\pb(E_{12})=1-K\pb(Y_1 < (1-\gamma)Y_2 )\geq 1-K\pb(Y_1 \leq (1-\gamma)Y_2.
\end{align}

We have
\begin{align*}
    \pb(Y_1 \leq (1 - \gamma)Y_2)=\pb((1 - \gamma)Y_2-Y_1\geq  0)=
    \pb \left(\sum_{i=1}^N \left( (1-\gamma) \hat R_{i, 2}-\hat R_{i,1}\right)\geq 0 \right)
\end{align*}

Let $V_i := (1-\gamma) \hat R_{i, 2}-\hat R_{i,1}$. Then $\{V_i\}$ is a sequence of bounded i.i.d. random variables, with   
\begin{equation*}
    V_i \in [-1, 1], \quad \EE\left[V_i\right]=(1-\gamma)\pi^R_2-\pi^R_1=-\frac{\gamma}{K}=-\frac{1-\alpha}{2K}.
\end{equation*}

Thus, using the Hoeffding inequality, we have that 
\begin{align}\label{eq:thm_2_proof_last_C2}
\pb(Y_1 \leq(1-\gamma)Y_2) &=\pb\p{ \sum_{i=1}^N V_i-N\EE\left[V_i\right]\geq-N\EE\left[V_i\right]} \nln
\leq & \exp\p{-\frac{2(\EE\left[V_i\right])^2N}{4}} \nln
\leq & \exp\p{-\frac{(1-\alpha)^2}{8K^2}N}. 
\end{align}
Combining \eqref{eq:thm_2_proof_last_C2} with \eqref{eq:proof_thm_2_C2}, we have that 
\begin{align}
    \pb_{C2}(\mbox{success}) \geq  1-K\exp\p{-\frac{(1-\alpha)^2}{8K^2}N}.
    \label{eq:C2_succ_iid}
\end{align}

Finally, combining \eqref{eq:C1_succ_iid} and \eqref{eq:C2_succ_iid} yields: 
\begin{equation*}
    p_+ \geq 1- \max \left\{ (K-1) \exp\p{-\frac{(1-\alpha)^2}{32K^2}N}, \, K\exp\p{-\frac{(1-\alpha)^2}{8K^2}N}  \right\}. 
\end{equation*}
This completes the proof of Theorem \ref{thm:pa_detection_simple}. 
\qed 

\subsection{Proof of Theorem \ref{thm:clt_pa_finite_time}}
\label{app:thm:clt_pa_finite_time}

\noindent \textbf{Theorem \ref{thm:clt_pa_finite_time}} (Central Limit Theorem for linear combinations of the Potential Action Statistic.)
    Fix $\lambda$ and $\mu_1,\ldots,\mu_K$ such that $\rho \in (0,1)$ and $\min_i \mu_i>0$. Fix ${\eta}=(\eta_1,\ldots,\eta_K)\in \mathbb{R}^K$ and let $h(\calQ)=(h_1(\calQ),\ldots,h_K(\calQ))$. Suppose the congestion control policy is JSQ. Then,  
    \begin{equation*}
        \sqrt{N}\p{\eta^T h(\calQ) -\eta^T  \pi^R } \Rightarrow \mathcal{N}(0, \sigma^2_{{\eta}}),
    \end{equation*}
    as $N\to \infty$, where $\sigma_{{\eta}}^2  :=  \var(\eta^T \hat R_{1})  + 2\sum_{n=2}^\infty \cov(\eta^T  \hat R_{1}, \eta^T \hat R_{n} ) < \infty$. \\

\emph{Proof outline.} There is an extensive literature on various approaches to establishing CLT in a Markov chain setting, and the steps that we will take in this proof are likely not unique. However, showing CLT in a countable space Markov chain is more delicate, and we have found these particular steps to be the easiest to verify given the nature of our setup. In particular, we will treat the queue length process as a continuous-time Markov chain over a countable state space. By analyzing the instantaneous drift of a Lyapunov function $V$ with respect to the generator of this Markov chain, we show that $Q(t)$ is $V$-uniformly ergodic (Theorem 7.1 of \citep{meyn1993stability}). Using Theorem 1(x) of \citep{gallegos2024equivalences} which equates different notions of geometric ergodicity, it follows $V$-uniform ergodicity implies that the chain is geometrically ergodic with respect to the total variation norm. Combining this latter notion of geometric ergodicity and the moment bounds on $h$ we obtain the CLT using Corollary 2 of \citep{jones2004markov}, attributed to \cite{ibragimov1962some}. 

\noindent \textbf{Proof.}  We start by showing that the queue length process is geometrically ergodic under JSQ, as isolated in the following result. Denote by $P^t(i,\cdot)$ the probability distribution of $Q(t)$ conditional on $Q(0) = i$ such that $P^t(i,A) := \pb(Q(t) \in A | Q(0) = i)$, $A \subset \zp^K$. 

\begin{theorem}[Geometric Ergodicity under JSQ]
    \label{theo:JSQ_Vergodic}
Suppose $\rho<1$. The congestion control policy is JSQ with a uniformly random tie-breaking rule. There exist $\beta <1$ and $c: \zp^K \to (0, \infty)$ such that 
\begin{equation}
    \lone{P^t(q, \cdot) - \pi }_{TV} \leq c(q) \beta^t, \quad t\in \rp,  q \in \zp^K, 
    \label{eq:geo_ergodic_qt}
\end{equation}
where $\lone{\cdot}_{TV}$ is the total variation norm, and $\pi$ is  $Q(\cdot)'s$ stationary distribution.
\end{theorem}

\begin{remark}
    Note that the geometric ergodicity of JSQ for the symmetric service rates case ($\mu_i$ is uniform) is known: it is typically proven using a straightforward argument involving stochastic comparison to an $M/M/1$ queue (cf.~\citep{luczak2006maximum}). However, this argument breaks down when the service rates are heterogeneous, such as in our setting with slower servers, because the servers are no longer exchangeable with one another. To the best of our knowledge, Theorem \ref{theo:JSQ_Vergodic} is the first to show the geometric ergodicity under JSQ with heterogeneous servers. 
\end{remark}
 
\emph{Proof of Theorem \ref{theo:JSQ_Vergodic}}:  Let $X(\cdot)$ be a continuous-time Markov chain taking value $\calX = \mathbb{Z}_+^d$, $d \in \mathbb{N}$. Let $G$ be the generator of $X$: 
$$G(i,j) := \lim_{t\downarrow 0 }\frac{1}{t} \p{\pb(X(t) = j | X(0)=i) - \mathbb{I}(i=j)}, \quad i, j \in \calX. $$
For a signed measure $\mu$ defined over $\calX$ and a positive function $f: \calX \to \mathbb{R}$, $f\geq 1$, the $f$-norm is defined as: $\lone{\pi}_f := \sup_{|g| \leq f }\int_\calX g(x) d\pi(x)$.  The following result is a crucial tool, which we re-state for the reader's convenience. 
\begin{theorem}[Theorems 6.1 \& 7.1, \citep{meyn1993stability}] 
\label{theo:drift_to_ergo}
Suppose $X(\cdot)$ is irreducible. If there exists a function $V: \calX \to \rp$ such that $V(x) \to \infty$ as $\lone{x} \to \infty$, satisfying for some $c, d \in (0, \infty)$: 
$$\Delta V(x) := \sum_{ x' \neq x } G(x,x')(V(x') - V(x)) \leq -c V(x) + d, \quad \forall x \in \calX,$$
then there exist $\beta <1$ and $B < \infty$ such that 
$$\lone{P^t(x, \cdot) - \pi}_f \leq Bf(x) \beta^t, \quad t\in \rp, x \in \calX, $$
where $f := V+1$ and  $\pi$ is the unique stationary distribution of $X(\cdot)$. 
\end{theorem}

We now construct a valid Lyapunov function for $Q(\cdot)$ and show it satisfies the drift condition in Theorem \ref{theo:drift_to_ergo}.  For $q \in \zp^K$, we will use the notation $q_{min}: = \min_i q_i$. Let $G$ be the generator for the Markov chain $Q(\cdot$). Then, $G(p,q')$ is nonzero only if $q$ and $q'$ differ only in one coordinate by 1. For all $q,q'$ where $|q_i - q_i'| = 1$ for some $i$ and $q_j= q_j'$ for all $j\neq i$, we have
\begin{equation}
 G(q,q') = \begin{cases}
      \lambda/|\{j: q_j = q_{min}\}|, & \text{if $q'_i= q_i + 1$ and $q_i = q_{min}$}, \\
      \mu_i, & \text{if $q'_i = q_i-1$ and $q_i >0$, }\\
      0, & \text{otherwise.}
    \end{cases}
    \label{eq:G_qt}
\end{equation}
where the first case corresponds to arrivals and the fact that the routing policy is JSQ with random tie-breaking, and the second case corresponds to the departures. 

Fix $\theta >0$. The Lyapunov function $V$ is defined by: 
\begin{equation*}
    V(q) :=\sum_{i=1}^K e^{\theta q_i}, \quad q \in \zp^K. 
\end{equation*} 
Using \eqref{eq:G_qt}, the drift of $V$ with respect to $G$ is given by
\begin{align*}
    \Delta V(q)=&\sum_{ q' \neq q} G(q,q') \left(V({q}')-V({q}) \right) \nln
    =& \lambda\left(e^{\theta(q_{min}+1)}-e^{\theta q_{min}}\right)+\sum_{i=1}^K\mu_i\left(e^{\theta(q_{i}-1)}-e^{\theta q_{i} }\right) \mathbb{I}(q_i>0) \nln
    =&  \lambda e^{\theta q_{min}}\left(e^{\theta}-1\right)+\sum_{i=1}^K\mu_i e^{\theta q_i}\left(e^{-\theta}-1\right) \mathbb{I}(q_i>0)\cr
    =&(e^\theta -1)\left [\lambda e^{\theta q_{min}}-e^{-\theta}\sum_{i=1}^K \mu_ie^{\theta q_i} +e^{-\theta}\sum_{i=1}^K \mu_i \mathbb{I}(q_i=0) \right] \nln
    \overset{(a)}{\leq} &(e^\theta -1)\left [\lambda e^{\theta q_{min}}-e^{-\theta}\sum_{i=1}^K \mu_ie^{\theta q_i} +e^{-\theta}\sum_{i=1}^K \mu_i \right]\cr
    \overset{(b)}{\leq} & (e^\theta -1)\left [\lambda \frac{\sum_{i=1}^K \mu_ie^{\theta q_i}}{\sum_{i=1}^K \mu_i}-e^{-\theta}\sum_{i=1}^K \mu_ie^{\theta q_i} +e^{-\theta}\sum_{i=1}^K \mu_i\right]\cr
    =&(e^\theta -1)\left [- \left(e^{-\theta}-\rho\right)\sum_{i=1}^K \mu_ie^{\theta q_i}+e^{-\theta}\sum_{i=1}^K \mu_i\right] \nln
    \leq  & -(e^\theta -1)(e^{-\theta } - \rho) \mu_{min} V(q) + (e^\theta -1)e^{-\theta}\lambda /\rho, 
\end{align*}
where $\rho = \lambda / \sum_i \mu_i $ and $\mu_{min} = \min_i \mu_i >0 $. Step $(a)$ follows from the fact that $\theta >0$ and hence $e^\theta -1>0$, and $(b)$ from the fact that $e^{\theta q_{min}}$ is no greater than any convex combination of $\{e^{\theta q_i}\}_{i = 1, \ldots, K}$. Setting $\theta = (1/2)\ln(1/\rho)$, we have that $e^{-\theta} - \rho = \sqrt{\rho} - \rho >0$ and 
\begin{equation}
    \Delta V(q) \leq -c V(q) + d, \quad \forall q \in \zp^K, 
    \label{eq:neg_drive_qt}
\end{equation}
where the constants $c = (e^\theta -1)(e^{-\theta } - \rho) \mu_{min} $ and $d = (e^\theta -1)e^{-\theta}\lambda /\rho$ are both strictly positive. 

Equation \eqref{eq:neg_drive_qt} thus verifies the drift condition in Theorem \ref{theo:drift_to_ergo}, and we conclude that the Markov chain $Q(t)$ is $V$-uniformly ergodic with respect to the norm $\lone{\cdot}_{V+1}$. It is not difficult show that this further implies the geometric ergodicity of $Q$ (cf.~Theorem 1(x) of \citep{gallegos2024equivalences}). This completes the proof of Theorem \ref{theo:JSQ_Vergodic}. \qed

Returning to the proof of Theorem \ref{thm:clt_pa_finite_time}. The columns of $\calQ$ form a discretely sampled embedded chain of $Q(\cdot)$: 
$$Q[n] := Q(t_n) = Q(L(n-1)), \quad n = \in \mathbb{N}. $$
It follows immediately from \eqref{eq:geo_ergodic_qt} that the discrete chain is also geometrically ergodic: there exist $\beta <1$ and $c: \zp^K \to (0, \infty)$ such that 
\begin{equation}
    \label{eq:geo_ergodic_qt_discrete}
    \lone{\pb(Q[n]\in \cdot | Q[1] = q) - \pi }_{TV} \leq c(q) \beta^n, \quad n \geq 2, q \in \zp^K.  
\end{equation}

To conclude the proof, we will use the following Markov chain CLT using the geometric ergodicity of $Q[\cdot]$. The following result is adapted from Corollary 2 of \cite{jones2004markov}, which is further based on \cite{ibragimov1962some}. 
\begin{theorem}[Corollary 2, \citep{jones2004markov}]
\label{theo:markov_CLT_ergodic}
Let $X[\cdot]$ be a Harris recurrent Markov chain with stationary distribution $\pi$ with the initialization $X[1]\sim \pi$. Let $f: \calX \to \mathbb{R}$ be  a Borel function such that $\EE_{ \pi}\left[|f(X)|^{\delta+2}\right] <\infty$ for some $\delta>0$. Suppose that \eqref{eq:geo_ergodic_qt_discrete} holds. Then, as $N\to \infty$ 
    \begin{equation*}
        \sqrt{N}\p{\frac{1}{N}\sum_{n=1}^N f(X[n]) - \EE_{ \pi}\left[f(X)\right] } \Rightarrow \mathcal{N}(0, \sigma^2_f), 
    \end{equation*}
    where $\sigma_f^2  := \var_{\pi}(f(X)) + 2\sum_{n=2}^\infty \cov(f(X[1]), f(X[n]) ) < \infty$. 
\end{theorem}

 We now apply Theorem \ref{theo:markov_CLT_ergodic} to our setting by letting the function $f$ be the linear combination of the potential action mappings $\eta^T \hat R(\cdot) $ in \eqref{eq:PA_def}. Because the potential actions take values in $[0,1]$ and $\eta$ is a fixed vector, the moment condition $\EE_{ \pi}(|f(X)|^{\delta+2}) <\infty$ is trivially satisfied. This concludes the proof of Theorem \ref{thm:clt_pa_finite_time}.  \qed


\subsection{Proof of Theorem \ref{thm:normal_appox_reliability}}
\label{app:thm:normal_appox_reliability}

\noindent \textbf{Theorem \ref{thm:normal_appox_reliability}}.
Consider the problem setting in Definition \ref{def:known_slowdown}. Suppose the congestion control policy is JSQ with random tie-breaking. Fix the summary statistics to be that of potential action and the decision rule a minorizing relative threshold rule with 
\begin{equation*}
     \gamma = \frac{1-1.1\alpha}{2}. 
\end{equation*}
Then, for all $L >0$, $K \geq 3$, $\alpha < 0.9$ and   $\lambda \geq 0.95 (\alpha + K-1)$, the reliability under  normal approximation satisfies: 
\begin{align*}
    \bar p_+ = & \min_{C \in \{C1, C2\}} \pb_{C}(\mbox{success}) \nln
&\geq  1- \max\left\{  (K-1)\Phi\left( -  \sqrt{N} \frac{1-1.1\alpha}{2K  \sigma_{C1, 1 \to 2}}   \right) 
, K  \Phi\p{ -\sqrt{N}\frac{1-1.1\alpha}{2K \sigma_{C2,1\to 2}}   } \right \}, 
\end{align*}
where $\sigma_{C1, k\to j}$ and $\sigma_{C2, k\to j}$ represent the values of $\sigma_{k \to j}$ under hypotheses $C1$ and $C2$, respectively, in Definition \ref{def:known_slowdown}. \\

\noindent \textbf{Proof.} Suppose C1 is true, so that server 1 slows down to a rate of $\alpha$. We have from Lemma \ref{lemma:ub and lb on ratio} that in this case the routing probability to server 1 would be smaller than that of other servers, and 
$$\pi^R_1 / \pi^R_j  \leq \alpha \left ( \frac{1}{\rho-(1-\rho)\frac{\alpha}{K-1}}\right)$$ 
 $j \neq 1$. In particular, when $\rho \geq 0.95$ and $K \geq 3$, the above simplifies to: 
 \begin{equation}
     \pi^R_1 / \pi^R_j  \leq \alpha(0.95 - 0.025)^{-1} \leq 1.1 \alpha :=     \alpha'. 
     \label{eq:alphaprime_bound}
 \end{equation}
Note that  that $\alpha' < 1 $ whenever $\alpha < 0.9$.  Therefore, under the choice of $\gamma = \frac{1-\alpha'}{2} = \frac{1-1.1\alpha}{2}$, we have that $1-\gamma = (1+\alpha')/2 \in (\alpha', 1)$ as desired. 

For all $j \neq 1$
\begin{align}
    (1-\gamma) \pi^R_j - \pi^R_1 = \frac{1+\alpha'}{2}  \pi^R_j - \pi^R_1  \geq \frac{1+\alpha'}{2}  \pi^R_j - \alpha'\pi^R_j =  \frac{1-\alpha'}{2} \pi^R_j \geq \frac{1-\alpha'}{2K}, 
    \label{eq:rhsHkj_C1}
\end{align}
where the first inequality follows from \eqref{eq:alphaprime_bound} and the second from  $\pi^R_j > 1/K$ for $j\neq 1$. Recall from \eqref{eq:HkH_j_event}
\begin{equation*}
    \calE_{kj} := \{\bar H_{kj}   <  \sqrt{N}\left[  (1-\gamma) \pi^R_j - \pi^R_k\right] \},
\end{equation*}
where $\bar H_{kj} \overset{d}{=} \calN(0, \sigma^2_{C1,k\to j})$. We have that
\begin{align}
       \pb_{C1}(\mbox{success})  = & \pb\left( \cap_{j\neq 1}  {\calE}_{1 j} \right) 
       = 1-\pb\left( \cup_{j\neq 1} \overline {\calE}_{1 j} \right) 
       \nln
       \geq & 1- \sum_{j\neq 1}  \bar\Phi\left(\sqrt{N}\left[  (1-\gamma) \pi^R_j - \pi^R_{1} \right] / \sigma_{C1,1 \to j} \right) \nln 
       = & 1- (K-1) \bar \Phi\left(\sqrt{N}\left[  (1-\gamma) \pi^R_2 - \pi^R_{1} \right] / \sigma_{C1,1 \to 2} \right) 
       \nln
       \geq & 1- (K-1)\bar\Phi\left(\sqrt{N}   \frac{1-\alpha'}{2K  \sigma_{C1,1 \to 2}}  \right) \nln
       = & 1- (K-1)\Phi\left(-\sqrt{N}   \frac{1-\alpha'}{2K  \sigma_{C1,1 \to 2}}  \right), 
       \label{eq:C1normalsucc}
\end{align}
where the last inequality follows from \eqref{eq:rhsHkj_C1}. 

Now suppose  C2 is true. Then all servers are identical and   $\pi_j^H = 1/K$ for all $j$. We have
\begin{equation*}
    (1-\gamma)\pi^R_j - \pi^R_k = -\gamma\pi^R_j  = -\frac{1-\alpha'}{2 K }, \quad \forall j\neq k. 
\end{equation*}

The success event under C2 corresponds to the rule outputting $\emptyset$, which further corresponds to the event $\cap_k \cup_{j \neq k} \overline \calE_{kj}$. We have that
\begin{align}
    \pb_{C2}(\mbox{success}) = &   \pb\left(\cap_k \cup_{j \neq k} \overline \calE_{kj} \right) \nln
    =& 1-\pb\left(\cup_k \cap_{j \neq k}  \calE_{kj} \right) \nln
    \geq & 1-K\pb\left( \cap_{j \neq 1}  \calE_{1j} \right) \nln
    \geq & 1-K\pb\left(  \calE_{12} \right) \nln
    =& 1-K \Phi\left(\sqrt{N}\left[  (1-\gamma) \pi^R_2 - \pi^R_{1} \right] / \sigma_{C2,1 \to 2} \right)  \nln
    =& 1-K \Phi\left(-\sqrt{N}\frac{1-\alpha'}{2 K \sigma_{C2, 1 \to 2} }\right) 
    \label{eq:C2normalsuccalt}
\end{align}

 Combining \eqref{eq:C1normalsucc} and \eqref{eq:C2normalsuccalt}, we arrive the at the characterization for reliability: 
\begin{align*}
\bar p_+ = & \min_{C \in \{C1, C2\}} \pb_{C}(\mbox{success}) \nln
\geq & 1- \max\left\{  (K-1)\Phi\left( - \sqrt{N}   \frac{1-\alpha'}{2K  \sigma_{C1,1 \to 2}}  \right) 
, K  \Phi\p{ -\sqrt{N} \frac{1-\alpha'}{2K \sigma_{C2,1\to 2}}}   \right \}. 
\end{align*}
Substituting $1.1\alpha$ for $\alpha'$ completes the proof of Theorem \ref{thm:normal_appox_reliability}. 
\qed

\end{APPENDICES}

\end{document}